\documentclass[prb,preprint,onecolumn,showpacs,preprintnumbers,amsmath,amssymb,superscriptaddress,floatfix]{revtex4-2}
\usepackage[T1]{fontenc}
\usepackage[ansinew]{inputenc}
\usepackage{graphics}
\usepackage{dcolumn}
\usepackage{bm}
\usepackage{amsthm}
\usepackage{amsmath}
\usepackage{amssymb}
\usepackage{diagbox}
\usepackage{hyperref}
\usepackage{url}
\usepackage{xcolor}

\begin{document}

\title{Magnetic neutron scattering from spherical nanoparticles with N\'{e}el surface anisotropy: Analytical treatment}
 
\author{Michael P. Adams}\email[Electronic address: ]{michael.adams@uni.lu}
\affiliation{Department of Physics and Materials Science, University of Luxembourg, 162A~avenue de la Faiencerie, L-1511~Luxembourg, Grand Duchy of Luxembourg}

\author{Andreas Michels}\email[Electronic address: ]{andreas.michels@uni.lu}
\affiliation{Department of Physics and Materials Science, University of Luxembourg, 162A~avenue de la Faiencerie, L-1511~Luxembourg, Grand Duchy of Luxembourg}  
 
\author{Hamid Kachkachi}\email[Electronic address: ]{hamid.kachkachi@univ-perp.fr}
\affiliation{Universit\'{e} de Perpignan via Domitia, Laboratoire PROMES CNRS UPR8521, Rambla de la Thermodynamique, Tecnosud, F-66100~Perpignan, France}
 
%%%%%%%%%%%%%%%%%%%%%%%%%%%%%%%%%%%%%%%%%%%%%%%%%%%%%%%%%%%%%%%%%%%

\begin{abstract}
The magnetization profile and the related magnetic small-angle neutron scattering cross section of a single spherical nanoparticle with N\'{e}el surface anisotropy is analytically investigated. We employ a Hamiltonian that comprises the isotropic exchange interaction, an external magnetic field, a uniaxial magnetocrystalline anisotropy in the core of the particle, and the N\'{e}el anisotropy at the surface. Using a perturbation approach, the determination of the magnetization profile can be reduced to a Helmholtz equation with Neumann boundary condition, whose solution is represented by an infinite series in terms of spherical harmonics and spherical Bessel functions. From the resulting infinite series expansion, we analytically calculate the Fourier transform, which is algebraically related to the magnetic small-angle neutron scattering cross section. The approximate analytical solution is compared to the numerical solution using the Landau-Lifshitz equation, which accounts for the full nonlinearity of the problem.
\end{abstract}

\date{\today}

\maketitle

%%%%%%%%%%%%%%%%%%%%%%%%%%%%%%%%%%%%%%%%%%%%%%%%%%%%%%%%%%%%%%%%%%%
 
\section{Introduction}
\label{sec1}

Magnetic small-angle neutron scattering (SANS) is a powerful technique for investigating spin structures on the mesoscopic length scale ($\sim 1$$-$$100 \, \mathrm{nm}$) and inside the volume of magnetic materials~\cite{rmp2019,michelsbook}. Recent SANS studies on magnetic nanoparticles, in particular employing spin-polarized neutrons, unanimously demonstrate that their spin textures are highly complex and exhibit a variety of nonuniform, canted, or core-shell-type configurations (see, \textit{e.g.}\ Refs.~\cite{disch2012,kryckaprl2014,ijiri2014,guenther2014,maurer2014,dennis2015,grutter2017,oberdick2018,krycka2019,benderapl2019,bersweiler2019,zakutna2020,dirkreview2022} and references therein). The magnetic SANS data analysis largely relies on structural form-factor-models for the cross section, borrowed from nuclear SANS, which do not properly account for the existing spin inhomogeneity inside magnetic nanoparticles or nanomagnets (NM).

Progress in magnetic SANS theory~\cite{michels2013,michels2014jmmm,mettus2015,erokhin2015,metmi2015,metmi2016,michelsPRB2016,michelsdmi2019,mistonov2019,metlov2022} strongly suggests that for the analysis of experimental magnetic SANS data the spatial nanometer scale variation of the orientation and magnitude of the magnetization vector field must be taken into account, and that macrospin-based models---assuming a \textit{uniform} magnetization---are not adequate. The starting point for a proper analysis of the scattering problem is a micromagnetic continuum expression for the magnetic energy of the system. In the static case, this then leads to the so-called Brown's equations, a set of nonlinear partial differential equations for the magnetization along with complex boundary conditions on the surface of the magnet. From these equations the Fourier image and the magnetic SANS cross section may be obtained.

In this paper, we present an \textit{analytical} treatment of the magnetic SANS cross section of a spherical NM with N\'{e}el's surface anisotropy~\cite{nee54jpr}. The manuscript is organized as follows: In Section~\ref{sec2}, we calculate the real-space spin structure of the spherical NM using classical micromagnetic theory within the second-order perturbation approach. In Section~\ref{sec3}, we compute the three-dimensional Fourier transform of the real-space spin structure, which directly yields the magnetic neutron scattering cross section and the pair-distance distribution function. The analytical results are benchmarked by comparing them to numerical finite-difference simulations using the Landau-Lifshitz equation of motion. Finally, Section~\ref{sec5} summarizes the main findings of this study.

We also make reference to our numerical study~\cite{adamsjacnum2022}, where in contrast to the present analytical work the full nonlinearity of the problem is considered.

\section{Micromagnetic Theory}
\label{sec2}

In the static micromagnetic approach~\cite{brown}, the magnetic configuration of a system is described by the continuous magnetization vector field $\mathbf{M}(\mathbf{r})$, which is subject to a constant magnitude $\|\mathbf{M}(\mathbf{r})\|=M_0$. The saturation magnetization $M_0$ is only a function of temperature. The normalized magnetization vector field is then defined as
\begin{align}
\mathbf{m}(\mathbf{r}) = \mathbf{M}(\mathbf{r})/M_{0} 
= [m_x(\mathbf{r}), m_y(\mathbf{r}), m_z(\mathbf{r})] .
\end{align}
Our Hamiltonian for the NM includes the isotropic exchange interaction, the Zeeman energy, a uniaxial magnetic anisotropy for spins in the core and N\'{e}el's surface anisotropy for those on the surface. In the continuum approach, it reads:
\begin{align}
\mathcal{H} 
&=
 - A \sum_{\alpha\in\{x,y,z\}} \int_V m_{\alpha} \Delta m_{\alpha} \; d^3 r
\nonumber  
\\
&- M_0 \mathbf{B}_{0} \int_{V} 
 \mathbf{m} 
 \; d^3r 
- K_c
\int_{V} \left(\mathbf{m}\cdot\mathbf{e}_{\mathrm{A}}\right)^{2}
\; d^3r 
\nonumber 
\\
&+ A \sum_{\alpha\in\{x,y,z\}}  
\oint_{\partial V} m_\alpha \nabla m_\alpha \cdot \mathbf{n}\; d^2r \nonumber 
\\
& - \frac{K_{s}}{2}   \sum_{\alpha\in\{x,y,z\}}\oint_{\partial V} \left|n_{\alpha}\right|m_{\alpha}^{2} \;  d^{2}r 
\label{eq:ContinuumHamiltonian3}
\end{align}
where $A$ is the exchange-stiffness constant, $\nabla$ is the Del operator, $\Delta$ is the Laplace operator, $\mathbf{B}_0 = \mu_0 \mathbf{H}_0$ is a constant applied magnetic field, $K_c > 0$ denotes the uniaxial core anisotropy constant, $\mathbf{e}_{\mathrm{A}}$ is a unit vector specifying the arbitrary core anisotropy axis, $K_s > 0$ is the N\'{e}el surface anisotropy constant~\cite{nee54jpr} with
\begin{align}
\mathbf{n}=[ \sin\theta \cos\phi, \sin\theta \sin\phi, \cos\theta] \label{eq:SurfaceNormalVector}
\end{align}
being the surface normal to the boundary of the NM~\cite{garanin2003, kachkachi07j3m}. In \eqref{eq:ContinuumHamiltonian3}, the two surface integrals take into account the boundary conditions for the magnetization on the surface ($\partial V$) of the NM of volume $V$, which result from the exchange interaction and the N\'{e}el term.

For small deviations from the homogeneous magnetization state, a perturbation approach is applicable. Let $\mathbf{m}_0$ be the principal unit vector (average direction) associated with $\mathbf{m}(\mathbf{r})$ and let the vector function $\boldsymbol{\psi}(\mathbf{r})\perp\mathbf{m}_0$ describe the spin misalignment. One can then write:
\begin{equation}
\mathbf{m}(\mathbf{r})=\mathbf{m}_{0}\sqrt{1-\|\boldsymbol{\psi}(\mathbf{r})\|^{2}}+\boldsymbol{\psi}(\mathbf{r}), \quad \|\mathbf{m}(\mathbf{r})\| = 1.
\label{eq:ExactLinarization}
\end{equation}
Assuming that $\psi_{x}, \psi_{y}, \psi_{z}\ll1$, the following second-order Maclaurin expansion in $\boldsymbol{\psi}$ is used to find an approximate closed-form solution for $\mathbf{m}(\mathbf{r})$:
\begin{equation}
\mathbf{m}\left(\mathbf{r}\right)\cong\mathbf{m}_{0}+\boldsymbol{\psi}\left(\mathbf{r}\right)-\frac{1}{2}\|\boldsymbol{\psi}(\mathbf{r})\|^{2}\mathbf{m}_{0} , \label{eq:SecondOrderApproximation}
\end{equation}
where $\mathbf{m}_0$ is taken as a known constant vector in subsequent calculations. By choosing the orthonormal vector base~\cite{garkac09prb}
\begin{align}
\begin{cases}
\displaystyle\mathbf{g}_0 = \mathbf{m}_0,
%\label{eq:UnitVectorBase_g0}
\\
\displaystyle\mathbf{g}_1 = \frac{\displaystyle \mathbf{m}_0 \times \mathbf{e}_{\mathrm{A}} }{\displaystyle \|\mathbf{m}_0 \times \mathbf{e}_{\mathrm{A}} \|},
%\label{eq:UnitVectorBase_g1}
\\
\displaystyle\mathbf{g}_2 = \frac{\displaystyle (\mathbf{m}_0 \cdot \mathbf{e}_{\mathrm{A}}) \cdot \mathbf{m}_0 - \mathbf{e}_{\mathrm{A}}}{\displaystyle \|(\mathbf{m}_0 \cdot \mathbf{e}_{\mathrm{A}}) \cdot \mathbf{m}_0 - \mathbf{e}_{\mathrm{A}}\|},
\end{cases}\label{eq:UnitVectorBase_g2}
\end{align}
the parametrization 
\begin{align}
\boldsymbol{\psi}(\mathbf{r}) 
= 
\psi_1(\mathbf{r}) \mathbf{g}_1 
+
\psi_2(\mathbf{r}) \mathbf{g}_2 ,
\label{eq:Parametrization}
\end{align}
and by introducing the dimensionless coordinates $\boldsymbol{\xi} = \mathbf{r}/R$ (with $\xi = \|\boldsymbol{\xi}\|= r/R$), where $\mathbf{r}$ is the position vector,
\begin{align}
\boldsymbol{r}=[r  \sin\theta \cos\phi , r \sin\theta \sin\phi, r \cos\theta] ,
\end{align}
and $R$ denotes the radius of the NM, the minimization of the Hamiltonian \eqref{eq:ContinuumHamiltonian3} leads to the well-known Helmholtz equation with Neumann boundary conditions on the unit sphere~\cite{kachkachi07j3m,garanin2003}:
\begin{align}
[\Delta_{\boldsymbol{\xi}} - \kappa_{\beta}^2]\psi_{\beta} & = 0  \; , \quad \beta \in \{1,2\}
\label{eq:DecoupledHelmholtzEquation1}
\\
\left.\frac{d\psi_{\beta}}{d \xi }\right|_{\xi =1} &= \sum_{\alpha \in \{x,y,z\}}  
 \chi_{\alpha}^{\beta}     |n_\alpha|  
\label{eq:DecoupledHelmholtzEquation2},
\end{align}
where the constants are defined as:
\begin{align}
\kappa_1^2 & = \mathbf{m}_0 \cdot \mathbf{b}_{0} 
+
2 k_c (\mathbf{m}_0 \cdot \mathbf{e}_{\mathrm{A}})^2 ,
\label{eq:DecoupledHelmholtzEquationCoefficient1}
\\
\kappa_2^2 & = \mathbf{m}_0 \cdot \mathbf{b}_{0} 
+
2 k_c \left[2 (\mathbf{m}_0 \cdot \mathbf{e}_{\mathrm{A}})^2 - 1\right] ,
\label{eq:DecoupledHelmholtzEquationCoefficient2}
\\
\chi_{\alpha}^{\beta}& = k_s (\mathbf{m}_{0}\cdot\mathbf{e}_\alpha)(\mathbf{g}_\beta\cdot\mathbf{e}_\alpha),
\label{eq:DecoupledHelmholtzEquationCoefficient2a}
\end{align}
with the dimensionless quantities
\begin{align}
k_{c}& = \frac{R^2 K_{c} }{A },
&
k_{s}& = \frac{R K_{s} }{A },
&
\mathbf{b}_{0}&= \frac{ R^2M_0}{A} \mathbf{B}_{0}.
\label{eq:DecoupledHelmholtzEquationCoefficient3}
\end{align}
The $\mathbf{e}_\alpha$ (with $\alpha=x,y,z$) in \eqref{eq:DecoupledHelmholtzEquationCoefficient2a} denote the unit vectors of the Cartesian laboratory coordinate frame (in which $\mathbf{n}$ and $\mathbf{r}$ are defined). We emphasize that there are only two independent differential equations for $\boldsymbol{\psi}$, which is a consequence of the constraint $\|\mathbf{m}(\mathbf{r})\| = 1$.

In our graphical representations, we will frequently use the following values: $k_c=0.1$ and $k_s=3.0$, which (using $R=5\,\mathrm{nm}$ and $A=10^{-11}\,\mathrm{J/m}$) correspond to $K_c = 40 \,\mathrm{kJ/m^3}$ and $K_s = 6 \,\mathrm{mJ/m^2}$~\cite{BATLLE2022,ohandley}. For $M_0 = 1.7 \times 10^6 \,\mathrm{A/m}$, the relation between $b_0$ (dimensionless) and the external field is $B_0 = 4/17 b_0 \times 1\mathrm{T}$.

The fundamental solution of the homogeneous Helmholtz equation \eqref{eq:DecoupledHelmholtzEquation1} is well known~\cite{weber2003essential,hobson2006mathematical}. Its nonsingular part can be expressed in spherical coordinates as an infinite series in terms of spherical harmonics $Y_{\ell m}(\theta,\phi)$ and spherical Bessel functions of the first kind $j_n(\mathrm{i}\kappa_{\beta}\xi)$,
\begin{align}
\psi_{\beta} = \sum_{\ell=0}^{\infty} \sum_{m=-\ell}^{\ell} c_{\ell m}^{\beta} j_{\ell}(\mathrm{i}\kappa_{\beta}\xi) Y_{\ell m}(\theta,\phi) .
\label{eq:HelmholtzFundamentalSolution}
\end{align}
The imaginary number `$\mathrm{i}$' in the argument of the spherical Bessel function is due to the negative sign in the Helmholtz equation \eqref{eq:DecoupledHelmholtzEquation1}. The expansion coefficients $c_{\ell m}^{\beta}$ are obtained from the Neumann boundary condition \eqref{eq:DecoupledHelmholtzEquation2} using the method of least squares (see Appendix~\ref{sec:SolutionOfTheBoundaryValueProblem}). From there it is seen that the zero-order term with $\ell =0$ vanishes, which physically makes sense, since the spin misalignment in our model is caused by the N\'{e}el surface anisotropy and, thus, due to symmetry reasons there is no misalignment at the center of the NM, \textit{i.e.}\ $\psi_{\beta}(\xi = 0, \theta, \phi) \equiv 0$. By contrast, the largest spin misalignment is found at the boundary of the NM, \textit{i.e.}\ $\xi=1$. Further, we find that the coefficients $c_{\ell m}^{\beta}$ vanish in the case of odd $\ell$ and $m$, they are real-valued, and even with respect to the index $m$, \textit{i.e.}\ $c_{\ell m}^{\beta}= c_{\ell,- m}^{\beta}$. Taking these properties into account, one can conveniently express the solution in terms of the associated Legendre polynomials $P_{\ell}^{m}(\cos\theta)$ with $\ell=2\nu$ and $m=2\mu$ [note that we use the convention that $Y_{\ell m}(\theta, \phi) = N_{\ell m} P_{\ell}^{m}(\cos\theta) \mathrm{e}^{\mathrm{i}m\phi}$~(p.~378 (14.30.1) in~\cite{olver2010nist})]:
\begin{align}
\psi_{\beta}  &=
 \sum_{\nu=1}^{\infty} \sum_{\mu=0}^{\nu} a_{\nu \mu}^{\beta} \Upsilon_{\nu}(\kappa_{\beta}\xi) P_{2\nu}^{2 \mu}(\cos\theta) \cos(2 \mu\phi),  \label{eq:HelmholtzFundamentalSolutionSymmetry}
\end{align}
where we define (compare p.~624--626 in \cite{weber2003essential})
\begin{align}
\Upsilon_{\nu}(\tau) &=j_{2\nu}(\mathrm{i}\tau)=   \frac{\sqrt{\pi}}{2} \sum_{s=0}^{\infty} \frac{(-1)^{\nu} (\tau/2)^{2(s+\nu)}}{s!\Gamma(2\nu +s+3/2)},\label{eq:SphericalBesselUpsilon}
\end{align}
and the expansion coefficients are given by
\begin{align}
a_{\nu \mu}^{\beta} &= \frac{2k_s N_{2\nu, 2\mu} }{1 + \delta_{\mu,0}} \frac{\mathbf{g}_{\beta} \cdot \operatorname{diag}\left[I_{2\nu, 2\mu}^{x}, I_{2\nu,2 \mu}^{y} , I_{2\nu, 2\mu}^{z} \right] \cdot \mathbf{m}_0}{\kappa_\beta \Upsilon_{\nu}'(\kappa_\beta)} 
\label{eq:CoefficientsHelmholtzEquation}
\end{align}
with
\begin{align}
 I_{\ell m}^{\alpha} &=\int_{0}^{2\pi} \int_{0}^{\pi} Y_{\ell, m}^{*} (\theta, \phi) |n_\alpha| \sin\theta \; d\theta d\phi
 \label{eq:IntegralCoefficients}
\end{align}
and 
\begin{align}
N_{\ell m} &= \sqrt{\frac{2\ell + 1}{4\pi} \frac{(\ell-m)!}{(\ell +m)!}}.
\end{align}
In \eqref{eq:CoefficientsHelmholtzEquation}, $\delta_{\mu,0}$ is the Kronecker delta function, $\operatorname{diag}[...]$ denotes a $3 \times 3$ diagonal matrix, and $\Upsilon_{\nu}'(\tau)$ is the first-order derivative of \eqref{eq:SphericalBesselUpsilon} with respect to $\tau$. For some small values of $\ell$ and $m$, the exact solutions of the integrals $I_{\ell m}^{\alpha}$ are listed in Appendix~\ref{sec:AppendixIntegralSolutions}. 

From \eqref{eq:CoefficientsHelmholtzEquation} it is seen that the functions $\psi_{\beta}$ depend linearly on $k_s$, such that for $k_s=0$ the magnetization of the NM is homogeneous (as expected). Since we assume that $\psi_{x}, \psi_{y}, \psi_{z}\ll1$, it is clear that the validity of our solution is restricted to a finite range $0 \le k_s \le k_{s,\mathrm{max}}$. Taking only the terms with $\nu=1$ into account (corresponding to $\ell=2$), the remaining (second-order) expression reads:
\begin{align}
\psi_{\beta}(\xi, \theta, \phi) &\approx 
-\frac{15k_s }{32} \frac{\Upsilon_{1}(\kappa_{\beta}\xi)}{\kappa_{\beta}\Upsilon_{1}'(\kappa_{\beta})}
\mathbf{g}_{\beta} \cdot
\boldsymbol{\mathcal{V}}(\theta,\phi) \cdot
\mathbf{m}_0 ,
\label{eq:SecondOrderApproximationOfPsi}
\end{align}
where
\begin{align}
\boldsymbol{\mathcal{V}}(\theta,\phi) &= \operatorname{diag}
\begin{bmatrix}
\cos^2\theta - 1/3 - \sin^2\theta\cos(2\phi) \\
\cos^2\theta - 1/3  + \sin^2\theta\cos(2\phi) \\
-2(\cos^2\theta - 1/3)
\end{bmatrix} .\label{eq:SecondOrderApproximationOfPsiAngularMatrix}
\end{align}
A reasonable approximation for small $\kappa_{\beta}$ in \eqref{eq:SecondOrderApproximationOfPsi} is obtained by taking into account the first two terms in the infinite series \eqref{eq:SphericalBesselUpsilon} for $\Upsilon_{\nu}(\tau)$. This results in the following expression [compare \eqref{eq:SecondOrderApproximationOfPsi}]:
\begin{align}
 \frac{\Upsilon_{1}(\kappa_{\beta}\xi)}{\kappa_{\beta}\Upsilon_{1}'(\kappa_{\beta})} \approx 
\frac{1}{4} \frac{\kappa_{\beta}^2 \xi^4 + 14 \xi^2}{\kappa_{\beta}^2 + 7} .
\end{align}
In the limit $\kappa_{\beta}\rightarrow 0$, this expression reduces to a quadratic function in $\xi$
\begin{align}
\lim_{\kappa_{\beta}\rightarrow 0}\left\{\frac{\Upsilon_{1}(\kappa_{\beta}\xi)}{\kappa_{\beta}\Upsilon_{1}'(\kappa_{\beta})}\right\} = \frac{\xi^2}{2} .
\label{eq:SecondOrderApproximationOfPsi4Limit1}
\end{align}
The case of an infinite applied magnetic field $\mathbf{B}_0$, or of a strong uniaxial core anisotropy [compare \eqref{eq:DecoupledHelmholtzEquationCoefficient1} and \eqref{eq:DecoupledHelmholtzEquationCoefficient2}], corresponds to the limit
\begin{align}
\lim_{\kappa_{\beta}\rightarrow \infty}\left\{\frac{\Upsilon_{1}(\kappa_{\beta}\xi)}{\kappa_{\beta}\Upsilon_{1}'(\kappa_{\beta})}\right\} = 0
\label{eq:SecondOrderApproximationOfPsi4Limit2},
\end{align}
which recovers the expected result of zero spin misalignment. Note that the limit $\kappa_{\beta}\rightarrow \infty$ is only obtained using all terms of the infinite series \eqref{eq:SphericalBesselUpsilon}. 

Of particular interest is the behavior of $\psi_{\beta}$ as a function of the radius $R$ of the NM. Inspecting the Hamiltonian \eqref{eq:ContinuumHamiltonian3}, it becomes clear that the surface anisotropy energy scales as $R^2$, while the uniaxial core anisotropy energy scales as $R^3$. Since the core and the surface anisotropy act in opposite ways (trying to make the spin structure more homogeneous, respectively, more inhomogeneous), we see that an increasing radius $R$ corresponds to a decreasing $\psi_{\beta}$. This behavior reflects the NM's surface-area-to-volume ratio. With \eqref{eq:SecondOrderApproximationOfPsi} it is not possible to make any prediction in this regard, because until this point we did not include the principal unit vector $\mathbf{m}_0$ into the minimization of the Hamiltonian. Generally, $\mathbf{m}_0$ is a function of $k_s$, $k_c$, $\mathbf{b}_0$, and $\mathbf{e}_{\mathrm{A}}$.

\begin{figure}[tb!]
\centering
\resizebox{0.70\columnwidth}{!}{\includegraphics{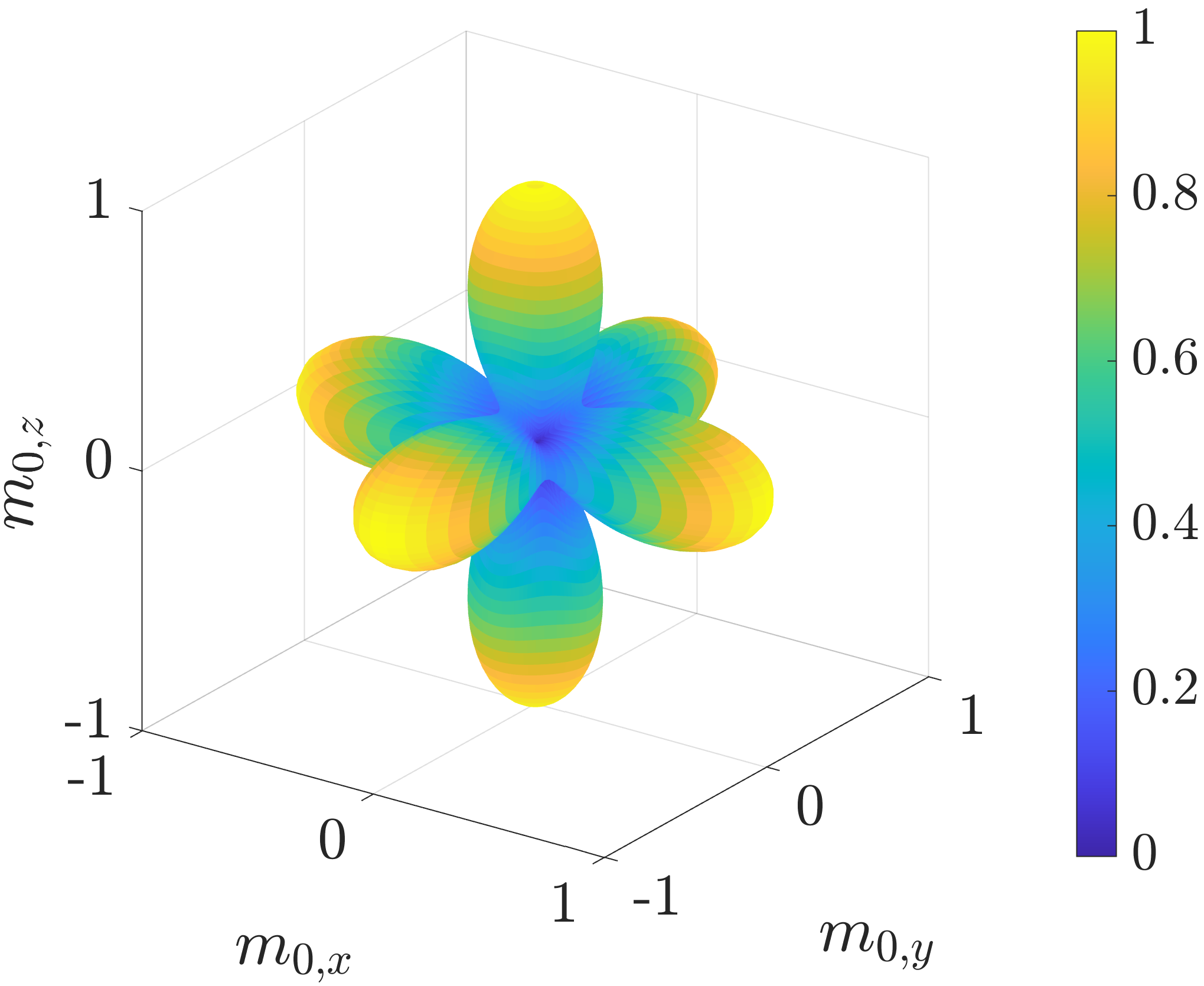}}
\caption{Normalized effective energy potential of the N\'{e}el surface anisotropy as a function of the Cartesian components of the average magnetization vector $\mathbf{m}_0 = [\sin\beta \cos\alpha, \sin\beta \sin\alpha, \cos\beta]$, computed via numerical integration of the surface contribution in \eqref{eq:ContinuumHamiltonian3} and by using the second-order approximation \eqref{eq:SecondOrderApproximationOfPsi}. Parameters are: $\mathbf{e}_{\mathrm{A}} = [0,0,1]$, $\mathbf{b}_0 = \mathbf{0}$, $k_c = 0.1$, and $k_s = 3.0$. The minima of the N\'{e}el surface contribution are in this case along the cubic space diagonals $\mathbf{m}_0 = [\pm 1 , \pm 1 , \pm 1]/\sqrt{3}$, while the maxima correspond to the Cartesian axes $\pm\mathbf{e}_x, \pm \mathbf{e}_y, \pm \mathbf{e}_z$. The effective energy potential has cubic symmetry and is approximately proportional to a function of the type $\simeq m_{0,x}^4 + m_{0,y}^4 + m_{0,z}^4$ (see also \cite{garanin2003}).}
\label{fig1}
\end{figure}

\begin{figure*}[tb!]
\centering
\resizebox{1.0\columnwidth}{!}{\includegraphics{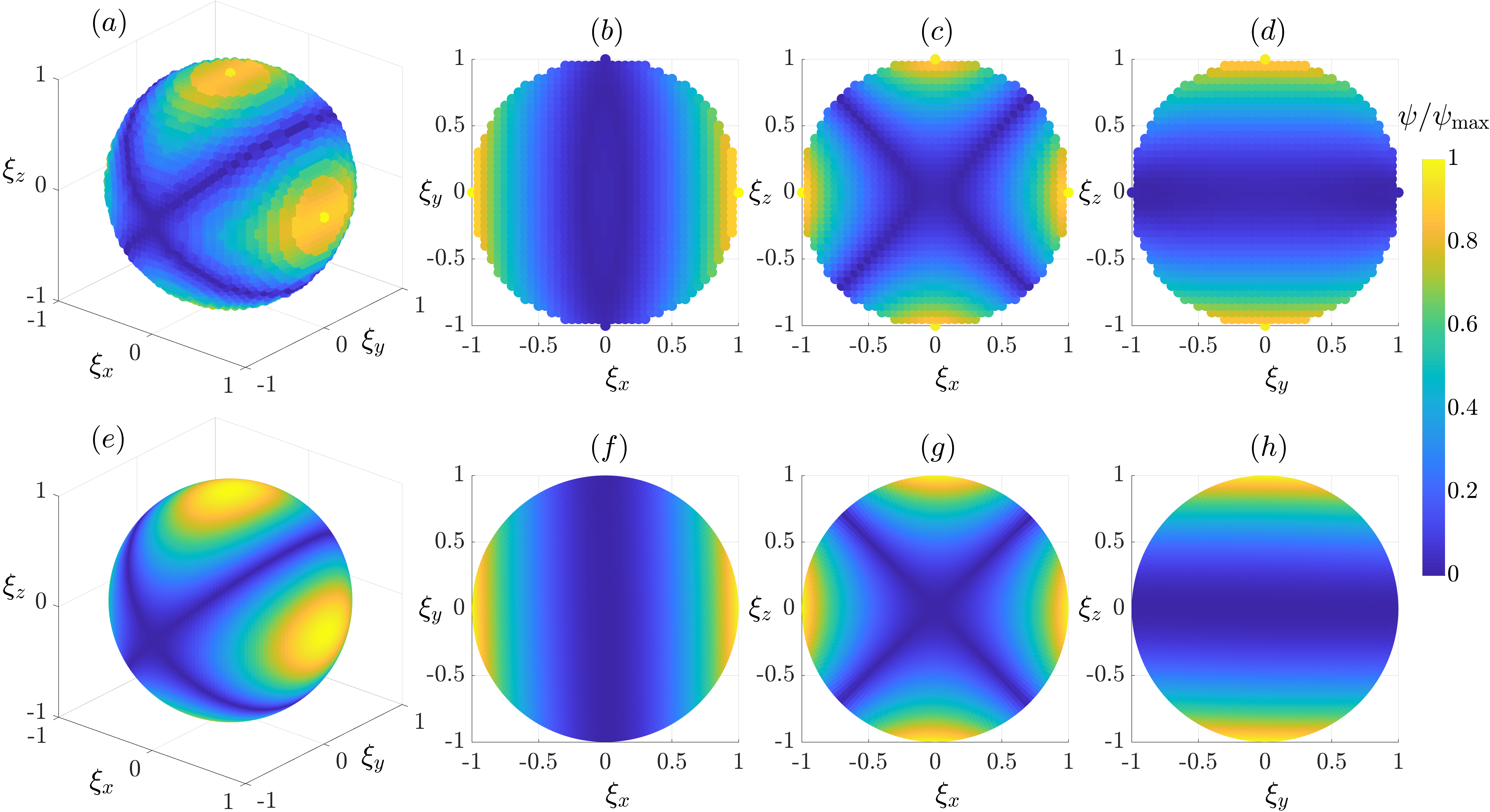}}
\caption{Comparison between the numerical solution using the Landau-Lifshitz equation (upper row) and the second-order analytical solution \eqref{eq:SecondOrderApproximationOfPsi} for $\|\boldsymbol{\psi}(\boldsymbol{\xi})\| = \sqrt{\psi_1^2(\boldsymbol{\xi}) + \psi_2^2(\boldsymbol{\xi})}$ (lower row). (\textit{a}) and (\textit{e}) show $\|\boldsymbol{\psi}\|$ on the boundary surface ($\xi=1$), while (\textit{b})$-$(\textit{d}) and (\textit{f})$-$(\textit{h}) display selected planar cuts in $(\xi_x,\xi_y,\xi_z)$~space. The following parameters are used: $\mathbf{e}_{\mathrm{A}} = \mathbf{e}_z$, $\mathbf{b}_0 = [0.4,0,0.4]$ ($B_0 \cong 133 \,\mathrm{mT}$), $k_c=0.1$, $k_s = 3.0$, and $\mathbf{m}_0 = [ \sin\beta \cos\alpha,  \sin\beta \sin\alpha, \cos\beta]$ where $\alpha = 0^{\circ}$ and $\beta = 40^{\circ}$.}
\label{fig2}
\end{figure*}

\begin{figure}[tb!]
\centering
\resizebox{0.50\columnwidth}{!}{\includegraphics{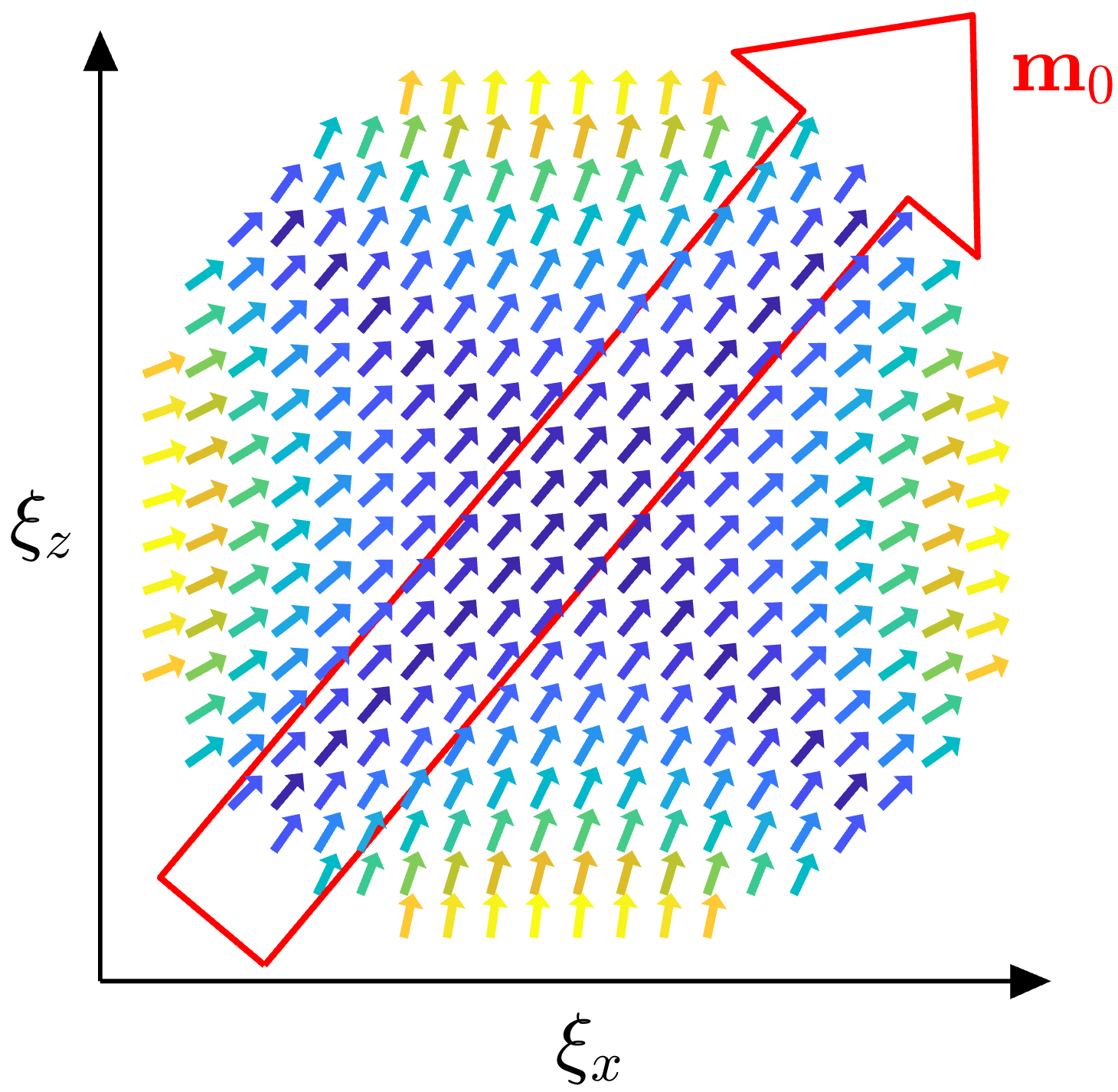}}
\caption{Real-space spin structure in the $\xi_x$-$\xi_z$~plane computed using \eqref{eq:ExactLinarization} and \eqref{eq:SecondOrderApproximationOfPsi}. Parameters are the same as in Fig.~\ref{fig2}. The external field $B_0 \cong 133 \,\mathrm{mT}$ is applied in the $\xi_x$-$\xi_z$~plane and inclined by an angle of $\beta = 40^{\circ}$ relative to the $\xi_z$~axis (compare to \cite{garanin2003}).}
\label{fig3}
\end{figure}

In the special case when the uniaxial anisotropy axis and the applied magnetic field are both directed parallel to the $z$~axis ($\mathbf{e}_{\mathrm{A}}=[0,0,1]$ and $\mathbf{b}_0=[0,0,b_0]$), the principal unit magnetization vector may be written as: 
\begin{align}
\mathbf{m}_0=[1/\sqrt{2}\sin\beta, 1/\sqrt{2}\sin\beta, \cos\beta] ,
\label{specialm0choice}
\end{align}
where $\beta \in [0, \arccos(1/\sqrt{3})]$. This choice is justified by the effective cubic symmetry of the N\'{e}el anisotropy as shown in Fig.~\ref{fig1}. This result was already predicted by Garanin and Kachkachi~\cite{garanin2003}. The solutions for $\psi_{1,2}(\xi,\theta,\phi)$ [using the particular $\mathbf{m}_0$ \eqref{specialm0choice}] then read:
\begin{align}
\psi_{1}  &\approx 
\frac{15k_s }{32} \frac{\Upsilon_{1}(\kappa_{1}\xi)}{\kappa_{1}\Upsilon_{1}'(\kappa_{1})}  \sin^2\theta \cos(2\phi)\sin\beta ,
\\
\psi_{2}  &\approx 
\frac{15k_s }{32} \frac{\Upsilon_{1}(\kappa_{2}\xi)}{\kappa_{2}\Upsilon_{1}'(\kappa_{2})} (1-3\cos^2\theta) \sin\beta\cos\beta .
\end{align}
In Fig.~\ref{fig2}, the analytical solution \eqref{eq:SecondOrderApproximationOfPsi} (lower row) is compared to the numerical solution based on the Landau-Lifshitz equation $\dot{\mathbf{m}} = -\gamma \mathbf{m}\times \mathbf{B}_{\mathrm{eff}} - \alpha \mathbf{m}\times(\mathbf{m}\times \mathbf{B}_{\mathrm{eff}})$ (upper row)~\cite{bertottibook}, where $\gamma$ is the gyromagnetic ratio, $\alpha$ is the damping constant, and the dot denotes the first-order time derivative (see our numerical study~\cite{adamsjacnum2022} for further details). Shown is the vector norm of the $\boldsymbol{\psi}(\boldsymbol{\xi})$~function scaled to its maximum value. From Fig.~\ref{fig2} it is seen that our analytical approximation is in qualitative agreement with the results from the numerical simulation. The corresponding real-space spin structure $\mathbf{m}(\boldsymbol{\xi})$ is displayed in Fig.~\ref{fig3}, where the surface spin disorder becomes clearly visible.

It is also instructive to compare our solution \eqref{eq:SecondOrderApproximationOfPsi} with that obtained using the Green function approach~\cite{garanin2003,kachkachi07j3m}. In particular, for $\xi$ located close to the surface, where the maximum spin misalignment with respect to $\mathbf{m}_0$ occurs, the Green function method yields the following approximate expression:
 \begin{align}
     \psi_{\beta}(\boldsymbol{\xi}) &\approx
     -\frac{15k_s}{32}\left[ 1 - \frac{\kappa_{\beta}^2}{14}\right] \xi^2
     \mathbf{g}_{\beta} \cdot
     \boldsymbol{\mathcal{V}}(\boldsymbol{\xi})
     \cdot \mathbf{m}_0, \label{eq:AnalyticsRealspaceSolution}
     \\
     \boldsymbol{\mathcal{V}}(\boldsymbol{\xi}) &=
     -\operatorname{diag}
     \begin{bmatrix}
     \xi_x^2/\xi^2 - 1/3  \\
     \xi_y^2/\xi^2 - 1/3  \\
     \xi_z^2/\xi^2 - 1/3
     \end{bmatrix}.
 \end{align}
This expression is also found when \eqref{eq:SecondOrderApproximationOfPsi} is expanded in $\kappa_{\beta}$ at the surface of the NM ($\xi=1$). While the infinite series approach using spherical harmonics and spherical Bessel functions yields an exact solution of the Helmholtz equation, the Green's function approach provides an approximate explicit expression of $\psi_{\beta}$ in terms of the coefficients $\kappa_\beta$. Indeed, as was shown in \cite{kachkachi07j3m}, in the presence of core anisotropy, the Green function as the kernel of the Helmholtz equation is only obtained as a perturbative series in $\kappa_{\beta}$. As such, \eqref{eq:AnalyticsRealspaceSolution} is restricted to small values of $\kappa_{\beta}$, \textit{i.e.}\ assuming that the core anisotropy and applied magnetic field are much smaller than the exchange coupling. This is manifest in~\eqref{eq:AnalyticsRealspaceSolution} by the presence of the factor $ 1 - \kappa_{\beta}^2/14$ which implies that the contribution of spin misalignment may diverge for too large $\kappa_{\beta}$ (\textit{i.e.}\ for a strong field and/or large core anisotropy).

\section{Magnetic SANS cross section}
\label{sec3}

\begin{figure}[tb!]
\centering
\resizebox{0.70\columnwidth}{!}{\includegraphics{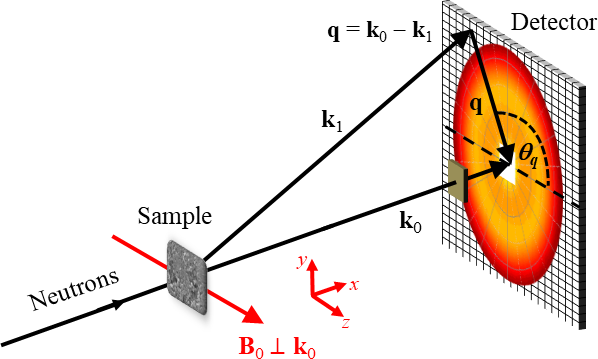}}
\caption{Sketch of the perpendicular scattering geometry ($\mathbf{B}_0 \perp \mathbf{k}_0$). The scattering vector $\mathbf{q}$ corresponds to the difference between the wave vectors of the incident ($\mathbf{k}_0$) and scattered ($\mathbf{k}_1$) neutrons. The angle $\theta_q$ specifies the orientation of $\mathbf{q}$ on the detector. In the small-angle approximation, the component of $\mathbf{q}$ along $\mathbf{k}_0$ is neglected.}
\label{fig4}
\end{figure}

The quantity of interest in experimental SANS studies is the elastic magnetic differential scattering cross section $d \Sigma_M / d \Omega$, which is usually recorded on a two-dimensional position-sensitive detector. For the most commonly used scattering geometry in magnetic SANS experiments, where the applied magnetic field $\mathbf{B}_0 \parallel \mathbf{e}_z$ is perpendicular to the wave vector $\mathbf{k}_0 \parallel \mathbf{e}_x$ of the incident neutrons (see Fig.~\ref{fig4}), $d \Sigma_M / d \Omega$ (for unpolarized neutrons) can be written as~\cite{rmp2019}:
\begin{eqnarray}
\frac{d \Sigma_M}{d \Omega}(\mathbf{q}) = \frac{8 \pi^3}{V} b_H^2 \left( |\widetilde{M}_x|^2 + |\widetilde{M}_y|^2 \cos^2 \theta_q \right. \nonumber \\ \left. + |\widetilde{M}_z|^2 \sin^2\theta_q - (\widetilde{M}_y \widetilde{M}_z^{\ast} + \widetilde{M}_y^{\ast} \widetilde{M}_z) \sin \theta_q \cos \theta_q \right) , \label{eq:equation1}
\end{eqnarray}
where $V$ is the scattering volume and $b_H = 2.91 \times 10^8 \, \mathrm{A}^{-1}\mathrm{m}^{-1}$ the magnetic scattering length in the small-angle regime (the atomic magnetic form factor is approximated by $1$, since we are dealing with forward scattering); $\widetilde{\mathbf{M}}(\mathbf{q}) = [ \widetilde{M}_x(\mathbf{q}), \widetilde{M}_y(\mathbf{q}), \widetilde{M}_z(\mathbf{q}) ]$ represents the magnetization vector field $\mathbf{M}(\mathbf{r})$ in Fourier space, $\theta_q$ denotes the angle between $\mathbf{q}$ and $\mathbf{B}_0$, and the asterisk `$*$' stands for the complex conjugate. Note that in the perpendicular scattering geometry, the Fourier components are evaluated in the plane $q_x = 0$.

The Fourier transform of the three-dimensional magnetization vector field (with a tilde above the symbol) is defined as follows
\begin{align}
\widetilde{\mathbf{M}}(\mathbf{q})&=  \frac{1}{(2\pi)^{3/2}} \int_V \mathbf{M}(\mathbf{r})  \exp\left(-\mathrm{i} \mathbf{q}\cdot\mathbf{r}\right) d^3r , \label{eq:FourierTransform_1}
\\
 \mathbf{M}(\mathbf{r})& = \frac{1}{(2\pi)^{3/2}} \int_V \widetilde{\mathbf{M}}(\mathbf{q})  \exp\left(\mathrm{i} \mathbf{q}\cdot\mathbf{r}\right) d^3q . \label{eq:InverseFourierTransform_1}
\end{align}
For subsequent calculations, we introduce the following dimensionless quantities
\begin{align}
\boldsymbol{\upsilon} &= \mathbf{q} R   , & 
 \widetilde{\boldsymbol{\mathcal{M}}} &= \frac{(2\pi)^{3/2} }{4\pi R^3 M_0} \widetilde{\mathbf{M}} , \label{eq:DimensionlessFourierTransform}
\end{align} 
and we express the dimensionless scattering vector in spherical coordinates as
\begin{align}
\boldsymbol{\upsilon} = [
\upsilon  \sin \theta_q \cos \phi_q,
\upsilon  \sin \theta_q \sin \phi_q ,
\upsilon \cos \theta_q ] . 
\end{align}
Next, in \eqref{eq:FourierTransform_1} we use the following first-order approximation for the real-space magnetization vector $\mathbf{m}(\boldsymbol{\xi})$ [see \eqref{eq:SecondOrderApproximation} and \eqref{eq:Parametrization}]
\begin{align}
\mathbf{m}(\boldsymbol{\xi}) =
 \mathbf{m}_0 
+
\sum_{\beta=1}^{2}\mathbf{g}_{\beta}  \psi_{\beta}(\boldsymbol{\xi}) .
\label{eq:FourierTransformFirstOrderApproximationRealSpace}
\end{align}
As shown in Appendix~\ref{sec:Derivation of the Fourier transform}, the final expression for the Fourier transform of the magnetization is then given by:
\begin{align}
&\widetilde{\boldsymbol{\mathcal{M}}}(\boldsymbol{\upsilon}) = 
\frac{j_1(\upsilon)}{\upsilon} \mathbf{m}_0 \nonumber \\
&+ 
\sum_{\beta=1}^{2}\mathbf{g}_{\beta} 
\sum_{\nu=1}^{\infty}\sum_{\mu=0}^{\nu}
(-1)^{\nu} a_{\nu \mu}^{\beta}    
\rho_{\nu}^{\beta}(\upsilon) P_{2\nu}^{2\mu}(\cos \theta_q) \cos(2\mu \phi_q),
\label{eq:ResultFourierTransform}
\end{align}
where
\begin{align}
\rho_{\nu}^{\beta}(\upsilon)=
-\frac{\upsilon j_{2\nu-1}(\upsilon) \Upsilon_{\nu}(\kappa_{\beta}) 
-  \kappa_{\beta} \mathcal{\aleph}_\nu(\kappa_{\beta})j_{2\nu}(\upsilon) }{\upsilon^2 + \kappa_{\beta}^2},
\label{eq:SphericalHankelTransform}
\end{align} 
\begin{align}
\mathcal{\aleph}_\nu (\tau) &=   
\frac{ \sqrt{\pi}}{2} 
\sum_{s=0}^{\infty} \frac{(-1)^{\nu}(\tau/2)^{2(s+\nu)-1}}{s!\Gamma(2\nu + s + 1/2)}    \label{eq:AlephBesselFunction},
\end{align}
and $\Upsilon_{\nu}(\kappa_{\beta})$ is given by \eqref{eq:SphericalBesselUpsilon}. The zero-order term $\propto j_1(\upsilon)/\upsilon$ in \eqref{eq:ResultFourierTransform} represents the form factor of a homogeneously magnetized sphere~\cite{michelsbook}. In the limiting case of an infinite applied magnetic field, which is equivalent to the limit $\kappa_{\beta} \rightarrow \infty$, the additional terms [second line in \eqref{eq:ResultFourierTransform}] vanish [compare \eqref{eq:SecondOrderApproximationOfPsi4Limit2}] and the spherical form factor remains. On the other hand, if $k_s=0$, the additional terms also vanish because from the physical point of view, the N\'{e}el surface anisotropy cancels and from \eqref{eq:CoefficientsHelmholtzEquation} we know that the coefficients $a_{\nu \mu}^{\beta}$ are linear in $k_s$. Taking only the terms with $\nu=1$ into account and setting $\phi_q = \pi/2$ ($\upsilon_x=0$), corresponding to the scattering geometry where the applied magnetic field $\mathbf{B}_0 \parallel \mathbf{e}_z$ is perpendicular to the wave vector $\mathbf{k}_0 \parallel \mathbf{e}_x$ of the incident neutrons (Fig.~\ref{fig4}), the expression for $\widetilde{\boldsymbol{\mathcal{M}}}(\boldsymbol{\upsilon})$ can be written as [compare with \eqref{eq:SecondOrderApproximationOfPsi}]:
\begin{align}
\widetilde{\boldsymbol{\mathcal{M}}}(\boldsymbol{\upsilon}) &\approx \frac{j_1(\upsilon)}{\upsilon} \mathbf{m}_0  \nonumber
\\
&- \frac{15 k_s}{32}
\sum_{\beta=1}^{2}  
\mathcal{R}_{\beta}(\upsilon) \left(\mathbf{g}_{\beta} \cdot \boldsymbol{\mathcal{V}}(\theta_q, \pi/2) \cdot \mathbf{m}_0 \right)\mathbf{g}_{\beta},
\label{eq:SecondOrderResultFourierTransform}
\end{align}
where the radial function is
\begin{align}
\mathcal{R}_{\beta}(\upsilon) &=
\frac{\upsilon j_{1}(\upsilon)}{\upsilon^2 + \kappa_{\beta}^2} 
\frac{\Upsilon_1(\kappa_{\beta})}{\kappa_{\beta} \Upsilon_1'(\kappa_{\beta})}
-
\frac{   j_{2}(\upsilon)  }{\upsilon^2 + \kappa_{\beta}^2} \frac{\aleph_1(\kappa_{\beta})}{\Upsilon_1'(\kappa_{\beta})} .
\end{align}
$\mathcal{R}_{\beta}(\upsilon)$ can be approximated for small $\kappa_{\beta}$ and when only terms up to $s=1$ in the infinite series \eqref{eq:SphericalBesselUpsilon} and \eqref{eq:AlephBesselFunction} are kept:
\begin{align}
\mathcal{R}_{\beta}(\upsilon) &=
\frac{1}{4} 
\frac{\upsilon j_{1}(\upsilon)}{\upsilon^2 + \kappa_{\beta}^2} 
\frac{\kappa_{\beta}^2  + 14}{\kappa_{\beta}^2 + 7}
-
\frac{7}{4}
\frac{   j_{2}(\upsilon)  }{\upsilon^2 + \kappa_{\beta}^2} 
\frac{\kappa_{\beta}^2 + 10}{\kappa_{\beta}^2 + 7} .
\end{align}
For small $\upsilon$-values, one finds the following limit:
\begin{align}
    \lim_{\upsilon\rightarrow 0} \mathcal{R}_{\beta}(\upsilon) = 0,
\end{align}
which is consistent with
\begin{align}
    \int_{V} \boldsymbol{\psi}(\mathbf{r}) \; d^3r = \boldsymbol{0} .
\end{align}
This can be seen by inspecting the definition of the Fourier transform  in \eqref{eq:FourierTransform_1}. Note that for $q \rightarrow 0$ the Fourier transform is proportional to the average of the magnetization vector field $\mathbf{M}$ and the maximum of this average is given by the homogeneous magnetization state. Using this result, the $\upsilon\rightarrow 0$ limit for the first-order approximation in $\boldsymbol{\psi}$ of the Fourier transform of the magnetization yields:
\begin{align}
    \lim_{\upsilon\rightarrow 0} \widetilde{\boldsymbol{\mathcal{M}}}(\upsilon,\theta_q,\phi_q) = \frac{1}{3} \mathbf{m}_0 .
\end{align}
Beyond the linear approximation in $\boldsymbol{\psi}$, a nonvanishing term appears in $\widetilde{\boldsymbol{\mathcal{M}}}$ in the limit $\upsilon\rightarrow 0$, which reduces the Fourier components relative to the homogeneous magnetization state. In the second order in $\boldsymbol{\psi}$, the result is [compare \eqref{eq:SecondOrderApproximation}]:
\begin{align}
    \lim_{\upsilon\rightarrow 0} \widetilde{\boldsymbol{\mathcal{M}}}(\upsilon,\theta_q,\phi_q) =\left[ \frac{1}{3}   - \frac{1}{2}  \int_{V} \|\boldsymbol{\psi}(\boldsymbol{\xi})\|^2\; d^3\xi\right] \cdot \mathbf{m}_0
\end{align}
Using \eqref{eq:DimensionlessFourierTransform} and 
\begin{align}
   \frac{d\Sigma_M}{d\Omega} = \frac{16\pi^2 R^6 M_0^2 b_H^2}{V}   \mathcal{S}_M,
\end{align}
the dimensionless two-dimensional magnetic SANS cross section $\mathcal{S}_M(\upsilon,\theta_q)$ can be straightforwardly obtained as [compare \eqref{eq:equation1}]
\begin{align}
    \mathcal{S}_M(\upsilon,\theta_q) &= 
    |\widetilde{\mathcal{M}}_x|^2 
    + |\widetilde{\mathcal{M}}_y|^2 \cos^2\theta_q 
    + |\widetilde{\mathcal{M}}_z|^2 \sin^2\theta_q \nonumber 
    \\   
     &- (\widetilde{\mathcal{M}}_y \widetilde{\mathcal{M}}_z^{\ast} + \widetilde{\mathcal{M}}_y^{\ast} \widetilde{\mathcal{M}}_z) \sin \theta_q \cos \theta_q .
     \label{eq:equation1dimless}
\end{align}
In limit $k_s \rightarrow 0$,  the resulting cross section from \eqref{eq:ResultFourierTransform} is
\begin{align}
    \lim_{k_s\rightarrow 0} \mathcal{S}_M(\upsilon,\theta_q) &=  \left(\frac{j_1(\upsilon)}{\upsilon}\right)^2   \left(m_{0,x}^2 + m_{0,y}^2 \cos^2 \theta_q    \right.  \nonumber
    \\
    &+  m_{0,z}^2 \sin^2  \theta_q
    - \left. 2 m_{0,y} m_{0,z} \sin \theta_q \cos \theta_q \right) .
    \label{eq:equation1dimlessks0}
\end{align}
The relation~\eqref{eq:equation1dimlessks0} nicely demonstrates that, depending on the orientation of the uniformly magnetized particle, different angular anisotropies become visible on the detector: For $\mathbf{m}_0 \parallel \mathbf{e}_x$ (\textit{i.e.}\ $m_{0y}=m_{0z}=0$), the scattering pattern is isotropic, while it exhibits a $\cos^2 \theta_q$ ($\sin^2 \theta_q$)~type shape when $\mathbf{m}_0 \parallel \mathbf{e}_y$ ($\mathbf{m}_0 \parallel \mathbf{e}_z$).

\begin{figure*}[tb!]
\centering
\resizebox{1.0\columnwidth}{!}{\includegraphics{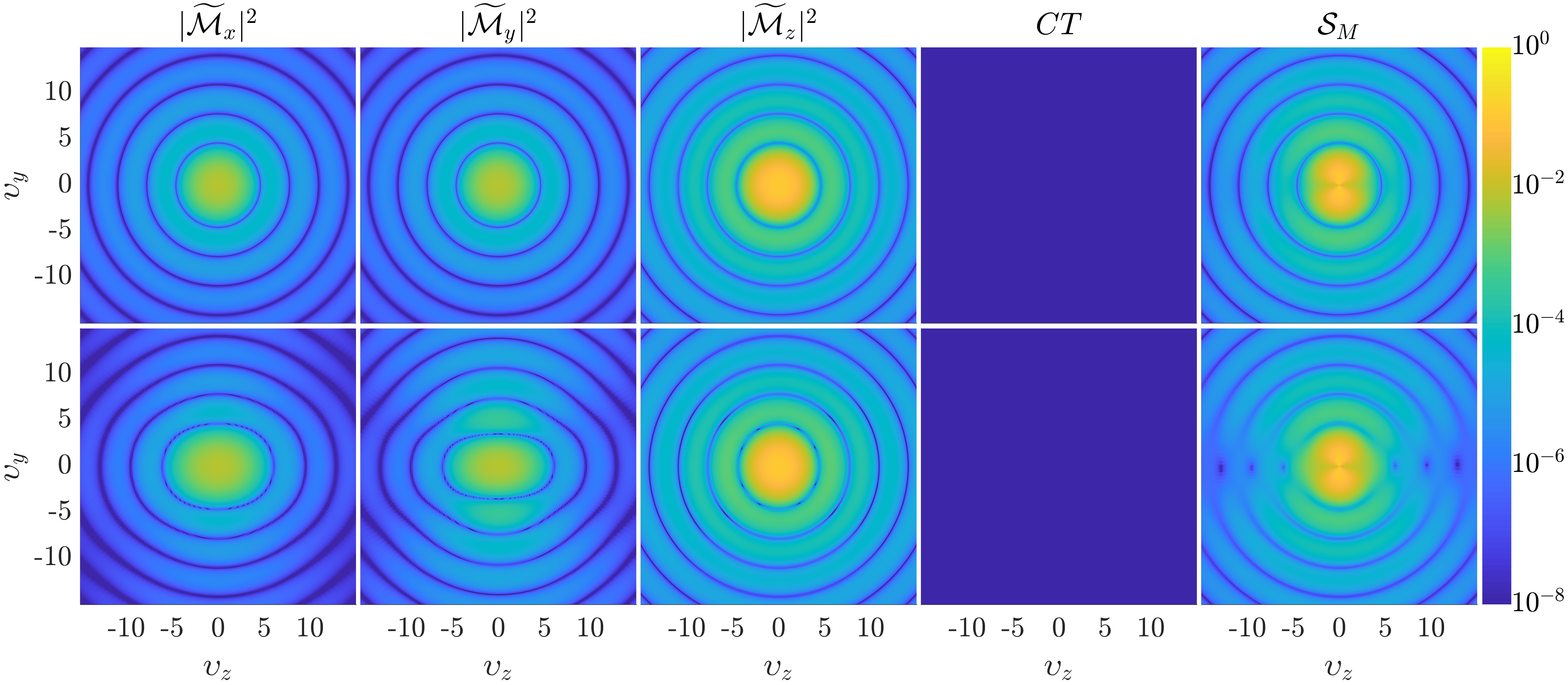}}
\caption{Results for the two-dimensional Fourier components $|\widetilde{\mathcal{M}}_x|^2$, $|\widetilde{\mathcal{M}}_y|^2$, $|\widetilde{\mathcal{M}}_z|^2$, $CT = -(\widetilde{\mathcal{M}}_y \widetilde{\mathcal{M}}_z^{\ast} + \widetilde{\mathcal{M}}_y^{\ast} \widetilde{\mathcal{M}}_z)$, and for the total magnetic SANS cross section $\mathcal{S}_M(\upsilon,\theta_q)$ [\eqref{eq:equation1dimless}] using expression \eqref{eq:SecondOrderResultFourierTransform}. The upper row shows the results taking into account only the zero-order term in \eqref{eq:SecondOrderResultFourierTransform}, which corresponds to the case of a homogeneously magnetized particle. The lower row displays the results when the second-order term ($\nu=1$) in \eqref{eq:SecondOrderResultFourierTransform} is taken into account. The parameters are: $\mathbf{e}_{\mathrm{A}} = \mathbf{e}_z$, $\mathbf{b}_0 = 0.1 \mathbf{e}_z$ [$B_0 \cong 24 \,\mathrm{mT}$], $k_c = 0.1$, $k_s=3$, and $\mathbf{m}_0 = [ \sin\beta \cos\alpha , \sin\beta \sin\alpha , \cos\beta]$. Since the N\'{e}el surface anisotropy has effectively a cubic symmetry (see Fig.~\ref{fig1}), we average $\mathcal{S}_M$ over the angles $\alpha=(45^{\circ}, 135^{\circ}, 225^{\circ}, 315^{\circ})$ and $\beta=20^{\circ}$. Logarithmic color scale is used.}
\label{fig5}
\end{figure*}

\begin{figure*}[tb!]
\centering
\resizebox{0.90\columnwidth}{!}{\includegraphics{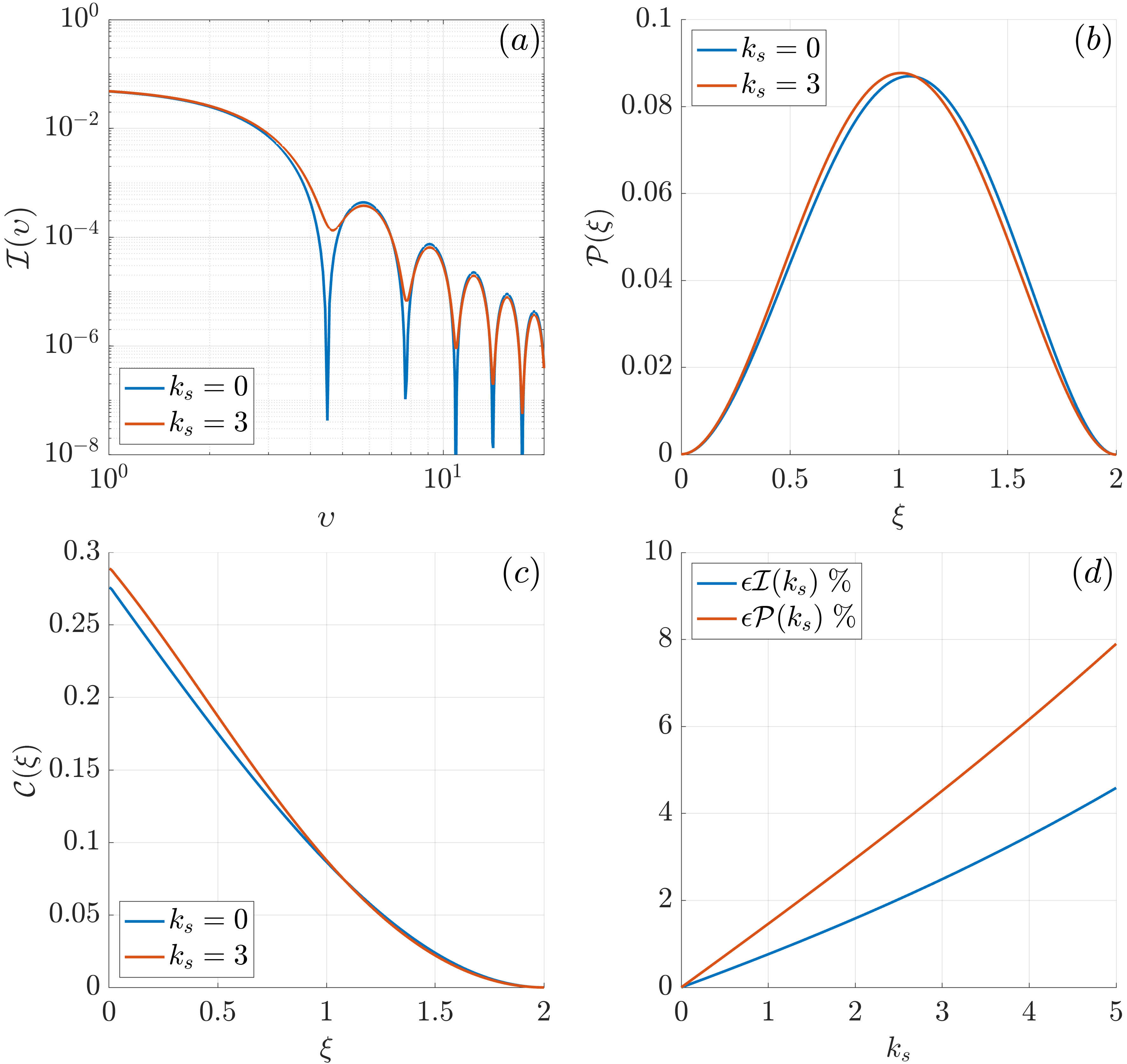}}
\caption{Results for (\textit{a})~the azimuthally-averaged SANS cross section $\mathcal{I}(\upsilon)$, (\textit{b})~the pair-distance distribution function $\mathcal{P}(\xi)$, (\textit{c})~the correlation function $\mathcal{C}(\xi)$, and (\textit{d}) the quantities $\epsilon\mathcal{I}(k_s)$ \eqref{eq:RelDevI} and $\epsilon\mathcal{P}(k_s)$ \eqref{eq:RelDevP} ($B_0 \cong 24 \,\mathrm{mT}$). For the homogeneous case (blue curves in (\textit{a}) - (\textit{c})), the surface anisotropy is set to $k_s=0$, and for the inhomogeneous case (red curves (\textit{a}) - (\textit{c})), we use the same parameters as in Fig.~\ref{fig5}. The functional dependence of $\mathcal{I}(\upsilon)$, $\mathcal{P}(\xi)$, and $\mathcal{C}(\xi)$ for the uniformly magnetized particle all correspond to the analytically well-known cases, \textit{i.e.}\ \eqref{eq:AzimuthalAverageks0}, \eqref{eq:AzimuthalAverageks1}, and \eqref{eq:AzimuthalAverageks2}.}
\label{fig6}
\end{figure*} 

Fig.~\ref{fig5} shows $\mathcal{S}_M(\upsilon,\theta_q)$ along with the contribution of the individual Fourier components to \eqref{eq:equation1dimless}. The upper row in Fig.~\ref{fig5} presents the result taking into account only the zero-order term ($j_1(\upsilon)/\upsilon \mathbf{m}_0$) from \eqref{eq:SecondOrderResultFourierTransform}, while in the lower row the second-order term ($\nu=1$) is additionally included. Since the zero-order term represents the case of a homogeneously magnetized NM, this comparison provides useful insights about the impact of the N\'{e}el surface anisotropy on the magnetic SANS cross section. In the case of a uniformly magnetized NM (upper row) the Fourier components $|\widetilde{\mathcal{M}}_x|^2$, $|\widetilde{\mathcal{M}}_y|^2$, and  $|\widetilde{\mathcal{M}}_z|^2$ are isotropic (rotational symmetry), while including the second-order terms (lower row) leads to an anisotropic behavior of the transverse components $|\widetilde{\mathcal{M}}_x|^2$ and $|\widetilde{\mathcal{M}}_y|^2$. The cross term ($CT$) averages to zero for both situations, and the dominating contribution to the magnetic SANS cross section is (for the parameters chosen in Fig.~\ref{fig5}) given by the $|\widetilde{\mathcal{M}}_z|^2$~component. Therefore, it may be concluded that the impact of the N\'{e}el surface anisotropy on $\mathcal{S}_M(\upsilon,\theta_q)$ is relatively small. By comparing the $\mathcal{S}_M(\upsilon,\theta_q)$ from the upper and the lower row, it is seen that by including the N\'{e}el surface anisotropy the circular symmetry of the zeros of $\mathcal{S}_M$ (deep blue colors) is broken. This feature becomes more clearly visible by analyzing the azimuthal average of $\mathcal{S}_M(\upsilon,\theta_q)$, which is readily computed as:
\begin{align}
    \mathcal{I}(\upsilon) = \frac{1}{2\pi} \int_{0}^{2\pi} \mathcal{S}_M(\upsilon,\theta_q) d \theta_q.
\end{align}
In the limit $k_s\rightarrow 0$, the azimuthal average corresponding to \eqref{eq:equation1dimlessks0} is:
\begin{align}
    \lim_{k_s\rightarrow 0} \mathcal{I}(\upsilon) = \left(\frac{j_1(\upsilon)}{\upsilon}\right)^2 \frac{\|\mathbf{m}_0\|^2 + (\mathbf{e}_x\cdot \mathbf{m}_0)^2}{2} .
    \label{eq:AzimuthalAverageks0} 
\end{align}
Moreover, we have also calculated the pair-distance distribution function,
\begin{align}
    \mathcal{P}(\xi) = \xi^2 \int_{0}^{\infty} \mathcal{I}(\upsilon) j_0(\upsilon \xi) \upsilon^2 d \upsilon ,
\end{align}
and the correlation function
\begin{align}
    \mathcal{C}(\xi) = \mathcal{P}(\xi)/\xi^2 .
\end{align}
In the limit $k_s\rightarrow 0$, the pair-distance distribution and the correlation function corresponding to \eqref{eq:AzimuthalAverageks0} are:
\begin{align}
    \lim_{k_s\rightarrow 0} \mathcal{P}(\xi) &= \frac{\pi \xi^2}{6}  \left(1 - \frac{3\xi}{4} + \frac{\xi^3}{16} \right) \frac{\|\mathbf{m}_0\|^2 + (\mathbf{e}_x\cdot \mathbf{m}_0)^2}{2} ,
    \label{eq:AzimuthalAverageks1} 
    \\
    \lim_{k_s\rightarrow 0} \mathcal{C}(\xi) &= \frac{\pi }{6}  \left(1 - \frac{3\xi}{4} + \frac{\xi^3}{16} \right) \frac{\|\mathbf{m}_0\|^2 + (\mathbf{e}_x\cdot \mathbf{m}_0)^2}{2} .
    \label{eq:AzimuthalAverageks2} 
\end{align}
These functions are graphically displayed in Fig.~\ref{fig6}. Due to the surface-anisotropy induced spin disorder, the form-factor maxima of $\mathcal{I}(\upsilon)$ [Fig.~\ref{fig6}(\textit{a})] are shifted to larger $q$~values (\textit{i.e.}\ smaller structures). Moreover, as already observed in numerical micromagnetic continuum simulations~\cite{laura2017,laura2020}, the oscillations are damped for the case of surface spin disorder, which mimics the effect of a particle-size distribution or of instrumental resolution. In agreement with this observation is the finding that the maximum of the $\mathcal{P}(\xi)$~function [Fig.~\ref{fig6}(\textit{b})] appears at smaller distances $\xi$ as compared to the homogeneous case. Likewise, due to spin disorder, the $\mathcal{C}(\xi)$~function [Fig.~\ref{fig6}(\textit{c})] exhibits a larger amplitude~\cite{mettus2015}.

To analyze the role of the surface anisotropy more quantitatively, we have computed the following quantities, which describe the deviation of the one-dimensional SANS cross section and of the pair-distance distribution function from the homogeneous particle case:
\begin{align}
    \epsilon\mathcal{I}(k_s) =\frac{\displaystyle\int_{0}^{\infty}|\mathcal{I}(k_s=0, \upsilon) - \mathcal{I}(k_s, \upsilon)|d \upsilon}{\displaystyle\int_{0}^{\infty}|\mathcal{I}(k_s=0, \upsilon)| d \upsilon} ,
    \label{eq:RelDevI}
    \\
    \epsilon\mathcal{P}(k_s) =\frac{\displaystyle\int_{0}^{2}|\mathcal{P}(k_s=0, \xi) - \mathcal{P}(k_s, \xi)|d \xi}{\displaystyle\int_{0}^{2}|\mathcal{P}(k_s=0, \xi)| d \xi} .
    \label{eq:RelDevP}
\end{align}
Fig.~\ref{fig6}(\textit{d}) depicts both $\epsilon\mathcal{I}(k_s)$ and $\epsilon\mathcal{P}(k_s)$ as a function of $k_s$. The difference is only of the order of a few percent, which suggests that the effect of surface anisotropy on the SANS observables is relatively weak within the present analytical approximation; see the numerical work~\cite{adamsjacnum2022}, which takes into account the full nonlinearity of the micromagnetic equations. However, this is only true for the magnetic interactions considered here. Taking into account the anisotropic and long-range dipole-dipole interaction and the asymmetric Dzyaloshinskii-Moriya interaction will very likely result in more inhomogeneous spin structures and in larger deviations from the macrospin model~\cite{laura2017,laura2020,hertel2021}. Likewise, for NM of elongated shapes, the surface anisotropy renders an additional first-order contribution to the effective energy \cite{garanin2003}, in addition to the second-order cubic contribution discussed above. This new shape-induced contribution could also lead to an enhancement of the spin misalignment. The analytical calculations presented here provide a general framework for future studies of more complicated (anisotropic) magnetic interactions.

\section{Conclusion}
\label{sec5}

We have analytically computed the magnetization distribution and the ensuing magnetic small-angle neutron scattering (SANS) cross section of a spherical nanoparticle. Our micromagnetic Hamiltonian takes into account the isotropic exchange interaction, an external magnetic field, a uniaxial anisotropy for the particle's core, and N\'{e}el anisotropy on its boundary. The resulting Helmholtz equation has been solved by expanding the real-space magnetization in terms of spherical Bessel functions and spherical harmonics. The central results are the infinite series \eqref{eq:HelmholtzFundamentalSolutionSymmetry} and its second-order expansion \eqref{eq:SecondOrderApproximationOfPsi} for the real-space magnetization, and the corresponding Fourier transforms \eqref{eq:ResultFourierTransform} and \eqref{eq:SecondOrderResultFourierTransform}. Using these expressions, the two-dimensional magnetic SANS cross section $\mathcal{S}_M(\upsilon,\theta_q)$, the azimuthally-averaged SANS signal $\mathcal{I}(\upsilon)$, and the correlation functions $\mathcal{P}(\xi)$ and $\mathcal{C}(\xi)$ were obtained and compared to the case of a homogeneous spin configuration (uniform magnetization vector field). The signature of N\'{e}el's surface anisotropy (of constant $k_s$) has been identified in all of these functions. However, its effect is relatively small, even for large values of $k_s$. Taking into account the magnetodipolar and/or the Dzyaloshinskii-Moriya interaction, or shape asymmetry, will likely result in configurations with stronger spin misalignment (\textit{e.g.}\ in vortex-type textures or skyrmions) and thereby in more prominent signatures in the SANS cross section and correlation function. These interactions are beyond the scope of the current analytical approach and will be considered in our future (numerical) works~\cite{adamsjacnum2022}.

\acknowledgments{Michael Adams and Andreas Michels thank the National Research Fund of Luxembourg for financial support (AFR Grant No.~15639149).}

%\bibliography{references}

\begin{thebibliography}{42}%
\makeatletter
\providecommand \@ifxundefined [1]{%
 \@ifx{#1\undefined}
}%
\providecommand \@ifnum [1]{%
 \ifnum #1\expandafter \@firstoftwo
 \else \expandafter \@secondoftwo
 \fi
}%
\providecommand \@ifx [1]{%
 \ifx #1\expandafter \@firstoftwo
 \else \expandafter \@secondoftwo
 \fi
}%
\providecommand \natexlab [1]{#1}%
\providecommand \enquote  [1]{``#1''}%
\providecommand \bibnamefont  [1]{#1}%
\providecommand \bibfnamefont [1]{#1}%
\providecommand \citenamefont [1]{#1}%
\providecommand \href@noop [0]{\@secondoftwo}%
\providecommand \href [0]{\begingroup \@sanitize@url \@href}%
\providecommand \@href[1]{\@@startlink{#1}\@@href}%
\providecommand \@@href[1]{\endgroup#1\@@endlink}%
\providecommand \@sanitize@url [0]{\catcode `\\12\catcode `\$12\catcode
  `\&12\catcode `\#12\catcode `\^12\catcode `\_12\catcode `\%12\relax}%
\providecommand \@@startlink[1]{}%
\providecommand \@@endlink[0]{}%
\providecommand \url  [0]{\begingroup\@sanitize@url \@url }%
\providecommand \@url [1]{\endgroup\@href {#1}{\urlprefix }}%
\providecommand \urlprefix  [0]{URL }%
\providecommand \Eprint [0]{\href }%
\providecommand \doibase [0]{https://doi.org/}%
\providecommand \selectlanguage [0]{\@gobble}%
\providecommand \bibinfo  [0]{\@secondoftwo}%
\providecommand \bibfield  [0]{\@secondoftwo}%
\providecommand \translation [1]{[#1]}%
\providecommand \BibitemOpen [0]{}%
\providecommand \bibitemStop [0]{}%
\providecommand \bibitemNoStop [0]{.\EOS\space}%
\providecommand \EOS [0]{\spacefactor3000\relax}%
\providecommand \BibitemShut  [1]{\csname bibitem#1\endcsname}%
\let\auto@bib@innerbib\@empty
%</preamble>
\bibitem [{\citenamefont {M\"uhlbauer}\ \emph {et~al.}(2019)\citenamefont
  {M\"uhlbauer}, \citenamefont {Honecker}, \citenamefont {P\'{e}rigo},
  \citenamefont {Bergner}, \citenamefont {Disch}, \citenamefont {Heinemann},
  \citenamefont {Erokhin}, \citenamefont {Berkov}, \citenamefont {Leighton},
  \citenamefont {Eskildsen},\ and\ \citenamefont {Michels}}]{rmp2019}%
  \BibitemOpen
  \bibfield  {author} {\bibinfo {author} {\bibfnamefont {S.}~\bibnamefont
  {M\"uhlbauer}}, \bibinfo {author} {\bibfnamefont {D.}~\bibnamefont
  {Honecker}}, \bibinfo {author} {\bibfnamefont {E.~A.}\ \bibnamefont
  {P\'{e}rigo}}, \bibinfo {author} {\bibfnamefont {F.}~\bibnamefont {Bergner}},
  \bibinfo {author} {\bibfnamefont {S.}~\bibnamefont {Disch}}, \bibinfo
  {author} {\bibfnamefont {A.}~\bibnamefont {Heinemann}}, \bibinfo {author}
  {\bibfnamefont {S.}~\bibnamefont {Erokhin}}, \bibinfo {author} {\bibfnamefont
  {D.}~\bibnamefont {Berkov}}, \bibinfo {author} {\bibfnamefont
  {C.}~\bibnamefont {Leighton}}, \bibinfo {author} {\bibfnamefont {M.~R.}\
  \bibnamefont {Eskildsen}},\ and\ \bibinfo {author} {\bibfnamefont
  {A.}~\bibnamefont {Michels}},\ }\href@noop {} {\bibfield  {journal} {\bibinfo
   {journal} {Rev. Mod. Phys.}\ }\textbf {\bibinfo {volume} {91}},\ \bibinfo
  {pages} {015004} (\bibinfo {year} {2019})}\BibitemShut {NoStop}%
\bibitem [{\citenamefont {Michels}(2021)}]{michelsbook}%
  \BibitemOpen
  \bibfield  {author} {\bibinfo {author} {\bibfnamefont {A.}~\bibnamefont
  {Michels}},\ }\href@noop {} {\emph {\bibinfo {title} {{Magnetic Small-Angle
  Neutron Scattering: A Probe for Mesoscale Magnetism Analysis}}}}\ (\bibinfo
  {publisher} {Oxford University Press},\ \bibinfo {address} {Oxford},\
  \bibinfo {year} {2021})\BibitemShut {NoStop}%
\bibitem [{\citenamefont {Disch}\ \emph {et~al.}(2012)\citenamefont {Disch},
  \citenamefont {Wetterskog}, \citenamefont {Hermann}, \citenamefont
  {Wiedenmann}, \citenamefont {Vainio}, \citenamefont {Salazar-Alvarez},
  \citenamefont {Bergstr\"om},\ and\ \citenamefont {Br\"uckel}}]{disch2012}%
  \BibitemOpen
  \bibfield  {author} {\bibinfo {author} {\bibfnamefont {S.}~\bibnamefont
  {Disch}}, \bibinfo {author} {\bibfnamefont {E.}~\bibnamefont {Wetterskog}},
  \bibinfo {author} {\bibfnamefont {R.~P.}\ \bibnamefont {Hermann}}, \bibinfo
  {author} {\bibfnamefont {A.}~\bibnamefont {Wiedenmann}}, \bibinfo {author}
  {\bibfnamefont {U.}~\bibnamefont {Vainio}}, \bibinfo {author} {\bibfnamefont
  {G.}~\bibnamefont {Salazar-Alvarez}}, \bibinfo {author} {\bibfnamefont
  {L.}~\bibnamefont {Bergstr\"om}},\ and\ \bibinfo {author} {\bibfnamefont
  {T.}~\bibnamefont {Br\"uckel}},\ }\href@noop {} {\bibfield  {journal}
  {\bibinfo  {journal} {New J. Phys.}\ }\textbf {\bibinfo {volume} {14}},\
  \bibinfo {pages} {013025} (\bibinfo {year} {2012})}\BibitemShut {NoStop}%
\bibitem [{\citenamefont {Krycka}\ \emph {et~al.}(2014)\citenamefont {Krycka},
  \citenamefont {Borchers}, \citenamefont {Booth}, \citenamefont {Ijiri},
  \citenamefont {Hasz}, \citenamefont {Rhyne},\ and\ \citenamefont
  {Majetich}}]{kryckaprl2014}%
  \BibitemOpen
  \bibfield  {author} {\bibinfo {author} {\bibfnamefont {K.~L.}\ \bibnamefont
  {Krycka}}, \bibinfo {author} {\bibfnamefont {J.~A.}\ \bibnamefont
  {Borchers}}, \bibinfo {author} {\bibfnamefont {R.~A.}\ \bibnamefont {Booth}},
  \bibinfo {author} {\bibfnamefont {Y.}~\bibnamefont {Ijiri}}, \bibinfo
  {author} {\bibfnamefont {K.}~\bibnamefont {Hasz}}, \bibinfo {author}
  {\bibfnamefont {J.~J.}\ \bibnamefont {Rhyne}},\ and\ \bibinfo {author}
  {\bibfnamefont {S.~A.}\ \bibnamefont {Majetich}},\ }\href@noop {} {\bibfield
  {journal} {\bibinfo  {journal} {Phys. Rev. Lett.}\ }\textbf {\bibinfo
  {volume} {113}},\ \bibinfo {pages} {147203} (\bibinfo {year}
  {2014})}\BibitemShut {NoStop}%
\bibitem [{\citenamefont {Hasz}\ \emph {et~al.}(2014)\citenamefont {Hasz},
  \citenamefont {Ijiri}, \citenamefont {Krycka}, \citenamefont {Borchers},
  \citenamefont {Booth}, \citenamefont {Oberdick},\ and\ \citenamefont
  {Majetich}}]{ijiri2014}%
  \BibitemOpen
  \bibfield  {author} {\bibinfo {author} {\bibfnamefont {K.}~\bibnamefont
  {Hasz}}, \bibinfo {author} {\bibfnamefont {Y.}~\bibnamefont {Ijiri}},
  \bibinfo {author} {\bibfnamefont {K.~L.}\ \bibnamefont {Krycka}}, \bibinfo
  {author} {\bibfnamefont {J.~A.}\ \bibnamefont {Borchers}}, \bibinfo {author}
  {\bibfnamefont {R.~A.}\ \bibnamefont {Booth}}, \bibinfo {author}
  {\bibfnamefont {S.}~\bibnamefont {Oberdick}},\ and\ \bibinfo {author}
  {\bibfnamefont {S.~A.}\ \bibnamefont {Majetich}},\ }\href@noop {} {\bibfield
  {journal} {\bibinfo  {journal} {Phys. Rev. B}\ }\textbf {\bibinfo {volume}
  {90}},\ \bibinfo {pages} {180405(R)} (\bibinfo {year} {2014})}\BibitemShut
  {NoStop}%
\bibitem [{\citenamefont {G\"unther}\ \emph {et~al.}(2014)\citenamefont
  {G\"unther}, \citenamefont {Honecker}, \citenamefont {Bick}, \citenamefont
  {Szary}, \citenamefont {Dewhurst}, \citenamefont {Keiderling}, \citenamefont
  {Feoktystov}, \citenamefont {Tsch\"ope}, \citenamefont {Birringer},\ and\
  \citenamefont {Michels}}]{guenther2014}%
  \BibitemOpen
  \bibfield  {author} {\bibinfo {author} {\bibfnamefont {A.}~\bibnamefont
  {G\"unther}}, \bibinfo {author} {\bibfnamefont {D.}~\bibnamefont {Honecker}},
  \bibinfo {author} {\bibfnamefont {J.-P.}\ \bibnamefont {Bick}}, \bibinfo
  {author} {\bibfnamefont {P.}~\bibnamefont {Szary}}, \bibinfo {author}
  {\bibfnamefont {C.~D.}\ \bibnamefont {Dewhurst}}, \bibinfo {author}
  {\bibfnamefont {U.}~\bibnamefont {Keiderling}}, \bibinfo {author}
  {\bibfnamefont {A.~V.}\ \bibnamefont {Feoktystov}}, \bibinfo {author}
  {\bibfnamefont {A.}~\bibnamefont {Tsch\"ope}}, \bibinfo {author}
  {\bibfnamefont {R.}~\bibnamefont {Birringer}},\ and\ \bibinfo {author}
  {\bibfnamefont {A.}~\bibnamefont {Michels}},\ }\href@noop {} {\bibfield
  {journal} {\bibinfo  {journal} {J. Appl. Cryst.}\ }\textbf {\bibinfo {volume}
  {47}},\ \bibinfo {pages} {992} (\bibinfo {year} {2014})}\BibitemShut
  {NoStop}%
\bibitem [{\citenamefont {Maurer}\ \emph {et~al.}(2014)\citenamefont {Maurer},
  \citenamefont {Gautrot}, \citenamefont {Ott}, \citenamefont {Chaboussant},
  \citenamefont {Zighem}, \citenamefont {Cagnon},\ and\ \citenamefont
  {Fruchart}}]{maurer2014}%
  \BibitemOpen
  \bibfield  {author} {\bibinfo {author} {\bibfnamefont {T.}~\bibnamefont
  {Maurer}}, \bibinfo {author} {\bibfnamefont {S.}~\bibnamefont {Gautrot}},
  \bibinfo {author} {\bibfnamefont {F.}~\bibnamefont {Ott}}, \bibinfo {author}
  {\bibfnamefont {G.}~\bibnamefont {Chaboussant}}, \bibinfo {author}
  {\bibfnamefont {F.}~\bibnamefont {Zighem}}, \bibinfo {author} {\bibfnamefont
  {L.}~\bibnamefont {Cagnon}},\ and\ \bibinfo {author} {\bibfnamefont
  {O.}~\bibnamefont {Fruchart}},\ }\href@noop {} {\bibfield  {journal}
  {\bibinfo  {journal} {Phys. Rev. B}\ }\textbf {\bibinfo {volume} {89}},\
  \bibinfo {pages} {184423} (\bibinfo {year} {2014})}\BibitemShut {NoStop}%
\bibitem [{\citenamefont {Dennis}\ \emph {et~al.}(2015)\citenamefont {Dennis},
  \citenamefont {Krycka}, \citenamefont {Borchers}, \citenamefont {Desautels},
  \citenamefont {van Lierop}, \citenamefont {Huls}, \citenamefont {Jackson},
  \citenamefont {Gruettner},\ and\ \citenamefont {Ivkov}}]{dennis2015}%
  \BibitemOpen
  \bibfield  {author} {\bibinfo {author} {\bibfnamefont {C.~L.}\ \bibnamefont
  {Dennis}}, \bibinfo {author} {\bibfnamefont {K.~L.}\ \bibnamefont {Krycka}},
  \bibinfo {author} {\bibfnamefont {J.~A.}\ \bibnamefont {Borchers}}, \bibinfo
  {author} {\bibfnamefont {R.~D.}\ \bibnamefont {Desautels}}, \bibinfo {author}
  {\bibfnamefont {J.}~\bibnamefont {van Lierop}}, \bibinfo {author}
  {\bibfnamefont {N.~F.}\ \bibnamefont {Huls}}, \bibinfo {author}
  {\bibfnamefont {A.~J.}\ \bibnamefont {Jackson}}, \bibinfo {author}
  {\bibfnamefont {C.}~\bibnamefont {Gruettner}},\ and\ \bibinfo {author}
  {\bibfnamefont {R.}~\bibnamefont {Ivkov}},\ }\href@noop {} {\bibfield
  {journal} {\bibinfo  {journal} {Adv. Funct. Mater.}\ }\textbf {\bibinfo
  {volume} {25}},\ \bibinfo {pages} {4300} (\bibinfo {year}
  {2015})}\BibitemShut {NoStop}%
\bibitem [{\citenamefont {Grutter}\ \emph {et~al.}(2017)\citenamefont
  {Grutter}, \citenamefont {Krycka}, \citenamefont {Tartakovskaya},
  \citenamefont {Borchers}, \citenamefont {Reddy}, \citenamefont {Ortega},
  \citenamefont {Ponce},\ and\ \citenamefont {Stadler}}]{grutter2017}%
  \BibitemOpen
  \bibfield  {author} {\bibinfo {author} {\bibfnamefont {A.~J.}\ \bibnamefont
  {Grutter}}, \bibinfo {author} {\bibfnamefont {K.~L.}\ \bibnamefont {Krycka}},
  \bibinfo {author} {\bibfnamefont {E.~V.}\ \bibnamefont {Tartakovskaya}},
  \bibinfo {author} {\bibfnamefont {J.~A.}\ \bibnamefont {Borchers}}, \bibinfo
  {author} {\bibfnamefont {K.~S.~M.}\ \bibnamefont {Reddy}}, \bibinfo {author}
  {\bibfnamefont {E.}~\bibnamefont {Ortega}}, \bibinfo {author} {\bibfnamefont
  {A.}~\bibnamefont {Ponce}},\ and\ \bibinfo {author} {\bibfnamefont
  {B.~J.~H.}\ \bibnamefont {Stadler}},\ }\href@noop {} {\bibfield  {journal}
  {\bibinfo  {journal} {ACS Nano}\ }\textbf {\bibinfo {volume} {11}},\ \bibinfo
  {pages} {8311} (\bibinfo {year} {2017})}\BibitemShut {NoStop}%
\bibitem [{\citenamefont {Oberdick}\ \emph {et~al.}(2018)\citenamefont
  {Oberdick}, \citenamefont {Abdelgawad}, \citenamefont {Moya}, \citenamefont
  {Mesbahi-Vasey}, \citenamefont {Kepaptsoglou}, \citenamefont {Lazarov},
  \citenamefont {Evans}, \citenamefont {Meilak}, \citenamefont {Skoropata},
  \citenamefont {van Lierop}, \citenamefont {Hunt-Isaak}, \citenamefont {Pan},
  \citenamefont {Ijiri}, \citenamefont {Krycka}, \citenamefont {Borchers},\
  and\ \citenamefont {Majetich}}]{oberdick2018}%
  \BibitemOpen
  \bibfield  {author} {\bibinfo {author} {\bibfnamefont {S.~D.}\ \bibnamefont
  {Oberdick}}, \bibinfo {author} {\bibfnamefont {A.}~\bibnamefont
  {Abdelgawad}}, \bibinfo {author} {\bibfnamefont {C.}~\bibnamefont {Moya}},
  \bibinfo {author} {\bibfnamefont {S.}~\bibnamefont {Mesbahi-Vasey}}, \bibinfo
  {author} {\bibfnamefont {D.}~\bibnamefont {Kepaptsoglou}}, \bibinfo {author}
  {\bibfnamefont {V.~K.}\ \bibnamefont {Lazarov}}, \bibinfo {author}
  {\bibfnamefont {R.~F.~L.}\ \bibnamefont {Evans}}, \bibinfo {author}
  {\bibfnamefont {D.}~\bibnamefont {Meilak}}, \bibinfo {author} {\bibfnamefont
  {E.}~\bibnamefont {Skoropata}}, \bibinfo {author} {\bibfnamefont
  {J.}~\bibnamefont {van Lierop}}, \bibinfo {author} {\bibfnamefont
  {I.}~\bibnamefont {Hunt-Isaak}}, \bibinfo {author} {\bibfnamefont
  {H.}~\bibnamefont {Pan}}, \bibinfo {author} {\bibfnamefont {Y.}~\bibnamefont
  {Ijiri}}, \bibinfo {author} {\bibfnamefont {K.~L.}\ \bibnamefont {Krycka}},
  \bibinfo {author} {\bibfnamefont {J.~A.}\ \bibnamefont {Borchers}},\ and\
  \bibinfo {author} {\bibfnamefont {S.~A.}\ \bibnamefont {Majetich}},\
  }\href@noop {} {\bibfield  {journal} {\bibinfo  {journal} {Sci. Rep.}\
  }\textbf {\bibinfo {volume} {8}},\ \bibinfo {pages} {3425} (\bibinfo {year}
  {2018})}\BibitemShut {NoStop}%
\bibitem [{\citenamefont {Ijiri}\ \emph {et~al.}(2019)\citenamefont {Ijiri},
  \citenamefont {Krycka}, \citenamefont {Hunt-Isaak}, \citenamefont {Pan},
  \citenamefont {Hsieh}, \citenamefont {Borchers}, \citenamefont {Rhyne},
  \citenamefont {Oberdick}, \citenamefont {Abdelgawad},\ and\ \citenamefont
  {Majetich}}]{krycka2019}%
  \BibitemOpen
  \bibfield  {author} {\bibinfo {author} {\bibfnamefont {Y.}~\bibnamefont
  {Ijiri}}, \bibinfo {author} {\bibfnamefont {K.~L.}\ \bibnamefont {Krycka}},
  \bibinfo {author} {\bibfnamefont {I.}~\bibnamefont {Hunt-Isaak}}, \bibinfo
  {author} {\bibfnamefont {H.}~\bibnamefont {Pan}}, \bibinfo {author}
  {\bibfnamefont {J.}~\bibnamefont {Hsieh}}, \bibinfo {author} {\bibfnamefont
  {J.~A.}\ \bibnamefont {Borchers}}, \bibinfo {author} {\bibfnamefont {J.~J.}\
  \bibnamefont {Rhyne}}, \bibinfo {author} {\bibfnamefont {S.~D.}\ \bibnamefont
  {Oberdick}}, \bibinfo {author} {\bibfnamefont {A.}~\bibnamefont
  {Abdelgawad}},\ and\ \bibinfo {author} {\bibfnamefont {S.~A.}\ \bibnamefont
  {Majetich}},\ }\href@noop {} {\bibfield  {journal} {\bibinfo  {journal}
  {Phys. Rev. B}\ }\textbf {\bibinfo {volume} {99}},\ \bibinfo {pages} {094421}
  (\bibinfo {year} {2019})}\BibitemShut {NoStop}%
\bibitem [{\citenamefont {Bender}\ \emph {et~al.}(2019)\citenamefont {Bender},
  \citenamefont {Honecker},\ and\ \citenamefont
  {Barqu\'{\i}n}}]{benderapl2019}%
  \BibitemOpen
  \bibfield  {author} {\bibinfo {author} {\bibfnamefont {P.}~\bibnamefont
  {Bender}}, \bibinfo {author} {\bibfnamefont {D.}~\bibnamefont {Honecker}},\
  and\ \bibinfo {author} {\bibfnamefont {L.~F.}\ \bibnamefont {Barqu\'{\i}n}},\
  }\href@noop {} {\bibfield  {journal} {\bibinfo  {journal} {Appl. Phys.
  Lett.}\ }\textbf {\bibinfo {volume} {115}},\ \bibinfo {pages} {132406}
  (\bibinfo {year} {2019})}\BibitemShut {NoStop}%
\bibitem [{\citenamefont {Bersweiler}\ \emph {et~al.}(2019)\citenamefont
  {Bersweiler}, \citenamefont {Bender}, \citenamefont {Vivas}, \citenamefont
  {Albino}, \citenamefont {Petrecca}, \citenamefont {M\"uhlbauer},
  \citenamefont {Erokhin}, \citenamefont {Berkov}, \citenamefont
  {Sangregorio},\ and\ \citenamefont {Michels}}]{bersweiler2019}%
  \BibitemOpen
  \bibfield  {author} {\bibinfo {author} {\bibfnamefont {M.}~\bibnamefont
  {Bersweiler}}, \bibinfo {author} {\bibfnamefont {P.}~\bibnamefont {Bender}},
  \bibinfo {author} {\bibfnamefont {L.~G.}\ \bibnamefont {Vivas}}, \bibinfo
  {author} {\bibfnamefont {M.}~\bibnamefont {Albino}}, \bibinfo {author}
  {\bibfnamefont {M.}~\bibnamefont {Petrecca}}, \bibinfo {author}
  {\bibfnamefont {S.}~\bibnamefont {M\"uhlbauer}}, \bibinfo {author}
  {\bibfnamefont {S.}~\bibnamefont {Erokhin}}, \bibinfo {author} {\bibfnamefont
  {D.}~\bibnamefont {Berkov}}, \bibinfo {author} {\bibfnamefont
  {C.}~\bibnamefont {Sangregorio}},\ and\ \bibinfo {author} {\bibfnamefont
  {A.}~\bibnamefont {Michels}},\ }\href@noop {} {\bibfield  {journal} {\bibinfo
   {journal} {Phys. Rev. B}\ }\textbf {\bibinfo {volume} {100}},\ \bibinfo
  {pages} {144434} (\bibinfo {year} {2019})}\BibitemShut {NoStop}%
\bibitem [{\citenamefont {Z\'akutn\'a}\ \emph {et~al.}(2020)\citenamefont
  {Z\'akutn\'a}, \citenamefont
  {Ni$\mathrm{\check{z}}$$\mathrm{\check{n}}$ansk\'y}, \citenamefont
  {Barnsley}, \citenamefont {Babcock}, \citenamefont {Salhi}, \citenamefont
  {Feoktystov}, \citenamefont {Honecker},\ and\ \citenamefont
  {Disch}}]{zakutna2020}%
  \BibitemOpen
  \bibfield  {author} {\bibinfo {author} {\bibfnamefont {D.}~\bibnamefont
  {Z\'akutn\'a}}, \bibinfo {author} {\bibfnamefont {D.}~\bibnamefont
  {Ni$\mathrm{\check{z}}$$\mathrm{\check{n}}$ansk\'y}}, \bibinfo {author}
  {\bibfnamefont {L.~C.}\ \bibnamefont {Barnsley}}, \bibinfo {author}
  {\bibfnamefont {E.}~\bibnamefont {Babcock}}, \bibinfo {author} {\bibfnamefont
  {Z.}~\bibnamefont {Salhi}}, \bibinfo {author} {\bibfnamefont
  {A.}~\bibnamefont {Feoktystov}}, \bibinfo {author} {\bibfnamefont
  {D.}~\bibnamefont {Honecker}},\ and\ \bibinfo {author} {\bibfnamefont
  {S.}~\bibnamefont {Disch}},\ }\href@noop {} {\bibfield  {journal} {\bibinfo
  {journal} {Phys. Rev. X}\ }\textbf {\bibinfo {volume} {10}},\ \bibinfo
  {pages} {031019} (\bibinfo {year} {2020})}\BibitemShut {NoStop}%
\bibitem [{\citenamefont {Honecker}\ \emph {et~al.}(2022)\citenamefont
  {Honecker}, \citenamefont {Bersweiler}, \citenamefont {Erokhin},
  \citenamefont {Berkov}, \citenamefont {Chesnel}, \citenamefont {Venero},
  \citenamefont {Qdemat}, \citenamefont {Disch}, \citenamefont {Jochum},
  \citenamefont {Michels},\ and\ \citenamefont {Bender}}]{dirkreview2022}%
  \BibitemOpen
  \bibfield  {author} {\bibinfo {author} {\bibfnamefont {D.}~\bibnamefont
  {Honecker}}, \bibinfo {author} {\bibfnamefont {M.}~\bibnamefont
  {Bersweiler}}, \bibinfo {author} {\bibfnamefont {S.}~\bibnamefont {Erokhin}},
  \bibinfo {author} {\bibfnamefont {D.}~\bibnamefont {Berkov}}, \bibinfo
  {author} {\bibfnamefont {K.}~\bibnamefont {Chesnel}}, \bibinfo {author}
  {\bibfnamefont {D.~A.}\ \bibnamefont {Venero}}, \bibinfo {author}
  {\bibfnamefont {A.}~\bibnamefont {Qdemat}}, \bibinfo {author} {\bibfnamefont
  {S.}~\bibnamefont {Disch}}, \bibinfo {author} {\bibfnamefont {J.~K.}\
  \bibnamefont {Jochum}}, \bibinfo {author} {\bibfnamefont {A.}~\bibnamefont
  {Michels}},\ and\ \bibinfo {author} {\bibfnamefont {P.}~\bibnamefont
  {Bender}},\ }\href@noop {} {\bibfield  {journal} {\bibinfo  {journal}
  {Nanoscale Adv.}\ }\textbf {\bibinfo {volume} {4}},\ \bibinfo {pages} {1026}
  (\bibinfo {year} {2022})}\BibitemShut {NoStop}%
\bibitem [{\citenamefont {Honecker}\ and\ \citenamefont
  {Michels}(2013)}]{michels2013}%
  \BibitemOpen
  \bibfield  {author} {\bibinfo {author} {\bibfnamefont {D.}~\bibnamefont
  {Honecker}}\ and\ \bibinfo {author} {\bibfnamefont {A.}~\bibnamefont
  {Michels}},\ }\href@noop {} {\bibfield  {journal} {\bibinfo  {journal} {Phys.
  Rev. B}\ }\textbf {\bibinfo {volume} {87}},\ \bibinfo {pages} {224426}
  (\bibinfo {year} {2013})}\BibitemShut {NoStop}%
\bibitem [{\citenamefont {Michels}\ \emph {et~al.}(2014)\citenamefont
  {Michels}, \citenamefont {Erokhin}, \citenamefont {Berkov},\ and\
  \citenamefont {Gorn}}]{michels2014jmmm}%
  \BibitemOpen
  \bibfield  {author} {\bibinfo {author} {\bibfnamefont {A.}~\bibnamefont
  {Michels}}, \bibinfo {author} {\bibfnamefont {S.}~\bibnamefont {Erokhin}},
  \bibinfo {author} {\bibfnamefont {D.}~\bibnamefont {Berkov}},\ and\ \bibinfo
  {author} {\bibfnamefont {N.}~\bibnamefont {Gorn}},\ }\href@noop {} {\bibfield
   {journal} {\bibinfo  {journal} {J. Magn. Magn. Mater.}\ }\textbf {\bibinfo
  {volume} {350}},\ \bibinfo {pages} {55} (\bibinfo {year} {2014})}\BibitemShut
  {NoStop}%
\bibitem [{\citenamefont {Mettus}\ and\ \citenamefont
  {Michels}(2015)}]{mettus2015}%
  \BibitemOpen
  \bibfield  {author} {\bibinfo {author} {\bibfnamefont {D.}~\bibnamefont
  {Mettus}}\ and\ \bibinfo {author} {\bibfnamefont {A.}~\bibnamefont
  {Michels}},\ }\href@noop {} {\bibfield  {journal} {\bibinfo  {journal} {J.
  Appl. Cryst.}\ }\textbf {\bibinfo {volume} {48}},\ \bibinfo {pages} {1437}
  (\bibinfo {year} {2015})}\BibitemShut {NoStop}%
\bibitem [{\citenamefont {Erokhin}\ \emph {et~al.}(2015)\citenamefont
  {Erokhin}, \citenamefont {Berkov},\ and\ \citenamefont
  {Michels}}]{erokhin2015}%
  \BibitemOpen
  \bibfield  {author} {\bibinfo {author} {\bibfnamefont {S.}~\bibnamefont
  {Erokhin}}, \bibinfo {author} {\bibfnamefont {D.}~\bibnamefont {Berkov}},\
  and\ \bibinfo {author} {\bibfnamefont {A.}~\bibnamefont {Michels}},\
  }\href@noop {} {\bibfield  {journal} {\bibinfo  {journal} {Phys. Rev. B}\
  }\textbf {\bibinfo {volume} {92}},\ \bibinfo {pages} {014427} (\bibinfo
  {year} {2015})}\BibitemShut {NoStop}%
\bibitem [{\citenamefont {Metlov}\ and\ \citenamefont
  {Michels}(2015)}]{metmi2015}%
  \BibitemOpen
  \bibfield  {author} {\bibinfo {author} {\bibfnamefont {K.~L.}\ \bibnamefont
  {Metlov}}\ and\ \bibinfo {author} {\bibfnamefont {A.}~\bibnamefont
  {Michels}},\ }\href@noop {} {\bibfield  {journal} {\bibinfo  {journal} {Phys.
  Rev. B}\ }\textbf {\bibinfo {volume} {91}},\ \bibinfo {pages} {054404}
  (\bibinfo {year} {2015})}\BibitemShut {NoStop}%
\bibitem [{\citenamefont {Metlov}\ and\ \citenamefont
  {Michels}(2016)}]{metmi2016}%
  \BibitemOpen
  \bibfield  {author} {\bibinfo {author} {\bibfnamefont {K.~L.}\ \bibnamefont
  {Metlov}}\ and\ \bibinfo {author} {\bibfnamefont {A.}~\bibnamefont
  {Michels}},\ }\href@noop {} {\bibfield  {journal} {\bibinfo  {journal} {Sci.
  Rep.}\ }\textbf {\bibinfo {volume} {6}},\ \bibinfo {pages} {25055} (\bibinfo
  {year} {2016})}\BibitemShut {NoStop}%
\bibitem [{\citenamefont {Michels}\ \emph {et~al.}(2016)\citenamefont
  {Michels}, \citenamefont {Mettus}, \citenamefont {Honecker},\ and\
  \citenamefont {Metlov}}]{michelsPRB2016}%
  \BibitemOpen
  \bibfield  {author} {\bibinfo {author} {\bibfnamefont {A.}~\bibnamefont
  {Michels}}, \bibinfo {author} {\bibfnamefont {D.}~\bibnamefont {Mettus}},
  \bibinfo {author} {\bibfnamefont {D.}~\bibnamefont {Honecker}},\ and\
  \bibinfo {author} {\bibfnamefont {K.~L.}\ \bibnamefont {Metlov}},\
  }\href@noop {} {\bibfield  {journal} {\bibinfo  {journal} {Phys. Rev. B}\
  }\textbf {\bibinfo {volume} {94}},\ \bibinfo {pages} {054424} (\bibinfo
  {year} {2016})}\BibitemShut {NoStop}%
\bibitem [{\citenamefont {Michels}\ \emph {et~al.}(2019)\citenamefont
  {Michels}, \citenamefont {Mettus}, \citenamefont {Titov}, \citenamefont
  {Malyeyev}, \citenamefont {Bersweiler}, \citenamefont {Bender}, \citenamefont
  {Peral}, \citenamefont {Birringer}, \citenamefont {Quan}, \citenamefont
  {Hautle}, \citenamefont {Kohlbrecher}, \citenamefont {Honecker},
  \citenamefont {Fern\'andez}, \citenamefont {Barqu\'{\i}n},\ and\
  \citenamefont {Metlov}}]{michelsdmi2019}%
  \BibitemOpen
  \bibfield  {author} {\bibinfo {author} {\bibfnamefont {A.}~\bibnamefont
  {Michels}}, \bibinfo {author} {\bibfnamefont {D.}~\bibnamefont {Mettus}},
  \bibinfo {author} {\bibfnamefont {I.}~\bibnamefont {Titov}}, \bibinfo
  {author} {\bibfnamefont {A.}~\bibnamefont {Malyeyev}}, \bibinfo {author}
  {\bibfnamefont {M.}~\bibnamefont {Bersweiler}}, \bibinfo {author}
  {\bibfnamefont {P.}~\bibnamefont {Bender}}, \bibinfo {author} {\bibfnamefont
  {I.}~\bibnamefont {Peral}}, \bibinfo {author} {\bibfnamefont
  {R.}~\bibnamefont {Birringer}}, \bibinfo {author} {\bibfnamefont
  {Y.}~\bibnamefont {Quan}}, \bibinfo {author} {\bibfnamefont {P.}~\bibnamefont
  {Hautle}}, \bibinfo {author} {\bibfnamefont {J.}~\bibnamefont {Kohlbrecher}},
  \bibinfo {author} {\bibfnamefont {D.}~\bibnamefont {Honecker}}, \bibinfo
  {author} {\bibfnamefont {J.~R.}\ \bibnamefont {Fern\'andez}}, \bibinfo
  {author} {\bibfnamefont {L.~F.}\ \bibnamefont {Barqu\'{\i}n}},\ and\ \bibinfo
  {author} {\bibfnamefont {K.~L.}\ \bibnamefont {Metlov}},\ }\href@noop {}
  {\bibfield  {journal} {\bibinfo  {journal} {Phys. Rev. B}\ }\textbf {\bibinfo
  {volume} {99}},\ \bibinfo {pages} {014416} (\bibinfo {year}
  {2019})}\BibitemShut {NoStop}%
\bibitem [{\citenamefont {Mistonov}\ \emph {et~al.}(2019)\citenamefont
  {Mistonov}, \citenamefont {Dubitskiy}, \citenamefont {Shishkin},
  \citenamefont {Grigoryeva}, \citenamefont {Heinemann}, \citenamefont
  {Sapoletova}, \citenamefont {Valkovskiy},\ and\ \citenamefont
  {Grigoriev}}]{mistonov2019}%
  \BibitemOpen
  \bibfield  {author} {\bibinfo {author} {\bibfnamefont {A.~A.}\ \bibnamefont
  {Mistonov}}, \bibinfo {author} {\bibfnamefont {I.~S.}\ \bibnamefont
  {Dubitskiy}}, \bibinfo {author} {\bibfnamefont {I.~S.}\ \bibnamefont
  {Shishkin}}, \bibinfo {author} {\bibfnamefont {N.~A.}\ \bibnamefont
  {Grigoryeva}}, \bibinfo {author} {\bibfnamefont {A.}~\bibnamefont
  {Heinemann}}, \bibinfo {author} {\bibfnamefont {N.~A.}\ \bibnamefont
  {Sapoletova}}, \bibinfo {author} {\bibfnamefont {G.~A.}\ \bibnamefont
  {Valkovskiy}},\ and\ \bibinfo {author} {\bibfnamefont {S.~V.}\ \bibnamefont
  {Grigoriev}},\ }\href@noop {} {\bibfield  {journal} {\bibinfo  {journal} {J.
  Magn. Magn. Mater.}\ }\textbf {\bibinfo {volume} {477}},\ \bibinfo {pages}
  {99} (\bibinfo {year} {2019})}\BibitemShut {NoStop}%
\bibitem [{\citenamefont {Zaporozhets}\ \emph {et~al.}(2022)\citenamefont
  {Zaporozhets}, \citenamefont {Oba}, \citenamefont {Michels},\ and\
  \citenamefont {Metlov}}]{metlov2022}%
  \BibitemOpen
  \bibfield  {author} {\bibinfo {author} {\bibfnamefont {V.~D.}\ \bibnamefont
  {Zaporozhets}}, \bibinfo {author} {\bibfnamefont {Y.}~\bibnamefont {Oba}},
  \bibinfo {author} {\bibfnamefont {A.}~\bibnamefont {Michels}},\ and\ \bibinfo
  {author} {\bibfnamefont {K.~L.}\ \bibnamefont {Metlov}},\ }\href@noop {}
  {\bibfield  {journal} {\bibinfo  {journal} {J. Appl. Cryst.}\ }\textbf
  {\bibinfo {volume} {xy}},\ \bibinfo {pages} {in press} (\bibinfo {year}
  {2022})}\BibitemShut {NoStop}%
\bibitem [{\citenamefont {N{\'e}el}(1954)}]{nee54jpr}%
  \BibitemOpen
  \bibfield  {author} {\bibinfo {author} {\bibfnamefont {L.}~\bibnamefont
  {N{\'e}el}},\ }\href@noop {} {\bibfield  {journal} {\bibinfo  {journal} {J.
  Phys. Radium}\ }\textbf {\bibinfo {volume} {15}},\ \bibinfo {pages} {225}
  (\bibinfo {year} {1954})}\BibitemShut {NoStop}%
\bibitem [{\citenamefont {Adams}\ \emph {et~al.}(2022)\citenamefont {Adams},
  \citenamefont {Michels},\ and\ \citenamefont {Kachkachi}}]{adamsjacnum2022}%
  \BibitemOpen
  \bibfield  {author} {\bibinfo {author} {\bibfnamefont {M.~P.}\ \bibnamefont
  {Adams}}, \bibinfo {author} {\bibfnamefont {A.}~\bibnamefont {Michels}},\
  and\ \bibinfo {author} {\bibfnamefont {H.}~\bibnamefont {Kachkachi}},\
  }\href@noop {} {\bibfield  {journal} {\bibinfo  {journal} {J. Appl. Cryst.}\
  }\textbf {\bibinfo {volume} {xy}},\ \bibinfo {pages} {abc} (\bibinfo {year}
  {2022})}\BibitemShut {NoStop}%
\bibitem [{\citenamefont {Brown~Jr.}(1963)}]{brown}%
  \BibitemOpen
  \bibfield  {author} {\bibinfo {author} {\bibfnamefont {W.~F.}\ \bibnamefont
  {Brown~Jr.}},\ }\href@noop {} {\emph {\bibinfo {title} {Micromagnetics}}}\
  (\bibinfo  {publisher} {Interscience Publishers},\ \bibinfo {address} {New
  York},\ \bibinfo {year} {1963})\BibitemShut {NoStop}%
\bibitem [{\citenamefont {Garanin}\ and\ \citenamefont
  {Kachkachi}(2003)}]{garanin2003}%
  \BibitemOpen
  \bibfield  {author} {\bibinfo {author} {\bibfnamefont {D.~A.}\ \bibnamefont
  {Garanin}}\ and\ \bibinfo {author} {\bibfnamefont {H.}~\bibnamefont
  {Kachkachi}},\ }\href@noop {} {\bibfield  {journal} {\bibinfo  {journal}
  {Phys. Rev. Lett.}\ }\textbf {\bibinfo {volume} {90}},\ \bibinfo {pages}
  {065504} (\bibinfo {year} {2003})}\BibitemShut {NoStop}%
\bibitem [{\citenamefont {{H. Kachkachi}}(2007)}]{kachkachi07j3m}%
  \BibitemOpen
  \bibfield  {author} {\bibinfo {author} {\bibnamefont {{H. Kachkachi}}},\
  }\href@noop {} {\bibfield  {journal} {\bibinfo  {journal} {J. Magn. Magn.
  Mater.}\ }\textbf {\bibinfo {volume} {316}},\ \bibinfo {pages} {248}
  (\bibinfo {year} {2007})}\BibitemShut {NoStop}%
\bibitem [{\citenamefont {Garanin}\ and\ \citenamefont
  {Kachkachi}(2009)}]{garkac09prb}%
  \BibitemOpen
  \bibfield  {author} {\bibinfo {author} {\bibfnamefont {D.~A.}\ \bibnamefont
  {Garanin}}\ and\ \bibinfo {author} {\bibfnamefont {H.}~\bibnamefont
  {Kachkachi}},\ }\href@noop {} {\bibfield  {journal} {\bibinfo  {journal}
  {Phys. Rev. B}\ }\textbf {\bibinfo {volume} {80}},\ \bibinfo {pages} {014420}
  (\bibinfo {year} {2009})}\BibitemShut {NoStop}%
\bibitem [{\citenamefont {Batlle}\ \emph {et~al.}(2022)\citenamefont {Batlle},
  \citenamefont {Moya}, \citenamefont {Escoda-Torroella}, \citenamefont
  {$\mathrm{\grave{O}}$. Iglesias}, \citenamefont {{Fraile Rodr{\'i}guez}},\
  and\ \citenamefont {Labarta}}]{BATLLE2022}%
  \BibitemOpen
  \bibfield  {author} {\bibinfo {author} {\bibfnamefont {X.}~\bibnamefont
  {Batlle}}, \bibinfo {author} {\bibfnamefont {C.}~\bibnamefont {Moya}},
  \bibinfo {author} {\bibfnamefont {M.}~\bibnamefont {Escoda-Torroella}},
  \bibinfo {author} {\bibnamefont {$\mathrm{\grave{O}}$. Iglesias}}, \bibinfo
  {author} {\bibfnamefont {A.}~\bibnamefont {{Fraile Rodr{\'i}guez}}},\ and\
  \bibinfo {author} {\bibfnamefont {A.}~\bibnamefont {Labarta}},\ }\href@noop
  {} {\bibfield  {journal} {\bibinfo  {journal} {J. Magn. Magn. Mater.}\
  }\textbf {\bibinfo {volume} {543}},\ \bibinfo {pages} {168594} (\bibinfo
  {year} {2022})}\BibitemShut {NoStop}%
\bibitem [{\citenamefont {O'Handley}(2000)}]{ohandley}%
  \BibitemOpen
  \bibfield  {author} {\bibinfo {author} {\bibfnamefont {R.~C.}\ \bibnamefont
  {O'Handley}},\ }\href@noop {} {\emph {\bibinfo {title} {Modern Magnetic
  Materials: Principles and Applications}}}\ (\bibinfo  {publisher} {Wiley},\
  \bibinfo {address} {New York},\ \bibinfo {year} {2000})\BibitemShut {NoStop}%
\bibitem [{\citenamefont {Weber}\ and\ \citenamefont
  {Arfken}(2003)}]{weber2003essential}%
  \BibitemOpen
  \bibfield  {author} {\bibinfo {author} {\bibfnamefont {H.~J.}\ \bibnamefont
  {Weber}}\ and\ \bibinfo {author} {\bibfnamefont {G.~B.}\ \bibnamefont
  {Arfken}},\ }\href@noop {} {\emph {\bibinfo {title} {{Essential Mathematical
  Methods for Physicists}}}}\ (\bibinfo  {publisher} {Academic Press},\
  \bibinfo {address} {Amsterdam},\ \bibinfo {year} {2003})\BibitemShut
  {NoStop}%
\bibitem [{\citenamefont {Riley}\ \emph {et~al.}(2006)\citenamefont {Riley},
  \citenamefont {Hobson},\ and\ \citenamefont
  {Bence}}]{hobson2006mathematical}%
  \BibitemOpen
  \bibfield  {author} {\bibinfo {author} {\bibfnamefont {K.~F.}\ \bibnamefont
  {Riley}}, \bibinfo {author} {\bibfnamefont {M.~P.}\ \bibnamefont {Hobson}},\
  and\ \bibinfo {author} {\bibfnamefont {S.~J.}\ \bibnamefont {Bence}},\
  }\href@noop {} {\emph {\bibinfo {title} {{Mathematical Methods for Physics
  and Engineering}}}}\ (\bibinfo  {publisher} {Cambridge University Press},\
  \bibinfo {address} {Cambridge},\ \bibinfo {year} {2006})\BibitemShut
  {NoStop}%
\bibitem [{\citenamefont {Olver}\ \emph {et~al.}(2010)\citenamefont {Olver},
  \citenamefont {Lozier}, \citenamefont {Boisvert},\ and\ \citenamefont
  {Clark}}]{olver2010nist}%
  \BibitemOpen
  \bibfield  {author} {\bibinfo {author} {\bibfnamefont {F.~W.~J.}\
  \bibnamefont {Olver}}, \bibinfo {author} {\bibfnamefont {D.~W.}\ \bibnamefont
  {Lozier}}, \bibinfo {author} {\bibfnamefont {R.~F.}\ \bibnamefont
  {Boisvert}},\ and\ \bibinfo {author} {\bibfnamefont {C.~W.}\ \bibnamefont
  {Clark}},\ }\href@noop {} {\emph {\bibinfo {title} {NIST Handbook of
  Mathematical Functions}}}\ (\bibinfo  {publisher} {Cambridge University
  Press},\ \bibinfo {address} {Cambridge},\ \bibinfo {year} {2010})\BibitemShut
  {NoStop}%
\bibitem [{\citenamefont {Bertotti}(1998)}]{bertottibook}%
  \BibitemOpen
  \bibfield  {author} {\bibinfo {author} {\bibfnamefont {G.}~\bibnamefont
  {Bertotti}},\ }\href@noop {} {\emph {\bibinfo {title} {Hysteresis in
  Magnetism}}}\ (\bibinfo  {publisher} {Academic Press},\ \bibinfo {address}
  {San Diego},\ \bibinfo {year} {1998})\BibitemShut {NoStop}%
\bibitem [{\citenamefont {Vivas}\ \emph {et~al.}(2017)\citenamefont {Vivas},
  \citenamefont {Yanes},\ and\ \citenamefont {Michels}}]{laura2017}%
  \BibitemOpen
  \bibfield  {author} {\bibinfo {author} {\bibfnamefont {L.~G.}\ \bibnamefont
  {Vivas}}, \bibinfo {author} {\bibfnamefont {R.}~\bibnamefont {Yanes}},\ and\
  \bibinfo {author} {\bibfnamefont {A.}~\bibnamefont {Michels}},\ }\href@noop
  {} {\bibfield  {journal} {\bibinfo  {journal} {Sci. Rep.}\ }\textbf {\bibinfo
  {volume} {7}},\ \bibinfo {pages} {13060} (\bibinfo {year}
  {2017})}\BibitemShut {NoStop}%
\bibitem [{\citenamefont {Vivas}\ \emph {et~al.}(2020)\citenamefont {Vivas},
  \citenamefont {Yanes}, \citenamefont {Berkov}, \citenamefont {Erokhin},
  \citenamefont {Bersweiler}, \citenamefont {Honecker}, \citenamefont
  {Bender},\ and\ \citenamefont {Michels}}]{laura2020}%
  \BibitemOpen
  \bibfield  {author} {\bibinfo {author} {\bibfnamefont {L.~G.}\ \bibnamefont
  {Vivas}}, \bibinfo {author} {\bibfnamefont {R.}~\bibnamefont {Yanes}},
  \bibinfo {author} {\bibfnamefont {D.}~\bibnamefont {Berkov}}, \bibinfo
  {author} {\bibfnamefont {S.}~\bibnamefont {Erokhin}}, \bibinfo {author}
  {\bibfnamefont {M.}~\bibnamefont {Bersweiler}}, \bibinfo {author}
  {\bibfnamefont {D.}~\bibnamefont {Honecker}}, \bibinfo {author}
  {\bibfnamefont {P.}~\bibnamefont {Bender}},\ and\ \bibinfo {author}
  {\bibfnamefont {A.}~\bibnamefont {Michels}},\ }\href@noop {} {\bibfield
  {journal} {\bibinfo  {journal} {Phys. Rev. Lett.}\ }\textbf {\bibinfo
  {volume} {125}},\ \bibinfo {pages} {117201} (\bibinfo {year}
  {2020})}\BibitemShut {NoStop}%
\bibitem [{\citenamefont {Pathak}\ and\ \citenamefont
  {Hertel}(2021)}]{hertel2021}%
  \BibitemOpen
  \bibfield  {author} {\bibinfo {author} {\bibfnamefont {S.~A.}\ \bibnamefont
  {Pathak}}\ and\ \bibinfo {author} {\bibfnamefont {R.}~\bibnamefont
  {Hertel}},\ }\href@noop {} {\bibfield  {journal} {\bibinfo  {journal} {Phys.
  Rev. B}\ }\textbf {\bibinfo {volume} {103}},\ \bibinfo {pages} {104414}
  (\bibinfo {year} {2021})}\BibitemShut {NoStop}%
\bibitem [{\citenamefont {Gradshteyn}\ and\ \citenamefont
  {Ryzhik}(2007)}]{gradshteyn2007table}%
  \BibitemOpen
  \bibfield  {author} {\bibinfo {author} {\bibfnamefont {I.~S.}\ \bibnamefont
  {Gradshteyn}}\ and\ \bibinfo {author} {\bibfnamefont {I.~M.}\ \bibnamefont
  {Ryzhik}},\ }\href@noop {} {\emph {\bibinfo {title} {Table of Integrals,
  Series, and Products}}},\ Vol.~\bibinfo {volume} {48}\ (\bibinfo  {publisher}
  {Elsevier/Academic Press, Amsterdam},\ \bibinfo {year} {2007})\BibitemShut
  {NoStop}%
\bibitem [{\citenamefont {Jackson}(1999)}]{jacksonen}%
  \BibitemOpen
  \bibfield  {author} {\bibinfo {author} {\bibfnamefont {J.~D.}\ \bibnamefont
  {Jackson}},\ }\href@noop {} {\emph {\bibinfo {title} {Classical
  Electrodynamics}}},\ \bibinfo {edition} {3rd}\ ed.\ (\bibinfo  {publisher}
  {Wiley},\ \bibinfo {address} {Hoboken},\ \bibinfo {year} {1999})\BibitemShut
  {NoStop}%
\end{thebibliography}


\appendix

\section{Solution of the Boundary Value Problem of the Helmholtz Equation}
\label{sec:SolutionOfTheBoundaryValueProblem}

The coefficients $c_{\ell m}^{\beta}$ in the fundamental solution \eqref{eq:HelmholtzFundamentalSolution} of the Helmholtz equation \eqref{eq:DecoupledHelmholtzEquation1} must be determined such that the Neumann boundary condition \eqref{eq:DecoupledHelmholtzEquation2} is satisfied. For this purpose, we use the method of least squares, where we make use of the orthogonality properties of the spherical harmonics $Y_{\ell m}(\theta, \phi)$. The normal derivative of \eqref{eq:HelmholtzFundamentalSolution} at the surface of the NM ($\xi=1$) is: 
\begin{align}
\left.\frac{d\psi_{\beta}}{d \xi}\right|_{\xi=1}= 
\sum_{\ell=0}^{\infty} \sum_{m=-\ell}^{\ell} c_{\ell m}^{\beta} [j_{\ell}( \mathrm{i} \kappa_{\beta}\xi)]_{\xi=1}' Y_{\ell m}(\theta, \phi),
\label{eq:AppxNormalDerivativeOfTheFundamentalSolution}
\end{align}
where
\begin{align}
[j_{\ell}( \mathrm{i} \kappa_{\beta}\xi)]_{\xi=1}' = \left.\frac{d}{d\xi} j_{\ell}( \mathrm{i} \kappa_{\beta}\xi)\right|_{\xi=1} .
\end{align}
Our goal is now to minimize the following error functional with respect to the coefficients $c_{\ell m}^{\beta}$:
\begin{align}
&\varepsilon[c_{\ell m}^{\beta} ] = \int_{0}^{2\pi}\int_{0}^{\pi} \left|
\sum_{\alpha \in \{x,y,z\}}  
 \chi_{\alpha}^{\beta}     |n_\alpha|    \right. \nonumber
\\
&-
\left.
\sum_{\ell=0}^{\infty} \sum_{m=-\ell}^{\ell} c_{\ell m}^{\beta} [j_{\ell}( \mathrm{i} \kappa_{\beta}\xi)]_{\xi=1}' Y_{\ell m}(\theta, \phi)
\right|^2 \sin\theta \; d\theta d\phi .
\label{eq:NeumannBoundaryConditionMSEAppx}
\end{align} 
The minimum of this error functional is found from the condition that the partial derivatives of $\varepsilon[c_{\ell m}^{\beta} ]$ with respect to the $(c_{ij}^{\beta})^{\ast}$ vanish:
\begin{align}
\frac{\partial \varepsilon[c_{\ell m}^{\beta}]}{\partial (c_{ij}^{\beta})^{\ast}} = 0 ,
\label{eq:AppxCLeastSquaresMinimization}
\end{align}
where `$*$' denotes the complex conjugate. Using the orthogonality relation \eqref{eq:OrthogonalityRelationSphericalHarmonicsAppxC} (p.~378 (14.30.8) in \cite{olver2010nist}) and by defining the integral $I_{ij}^{\alpha}$ by \eqref{eq:ResumingIntegralAppxC},
\begin{align}
\int_{0}^{2\pi}\int_{0}^{\pi} Y_{\ell m}(\theta, \phi)
 Y_{ij}^{*}(\theta, \phi)   \sin\theta \; d\theta d\phi &= \delta_{\ell i} \delta_{m j} ,
 \label{eq:OrthogonalityRelationSphericalHarmonicsAppxC}
 \\
 \int_{0}^{2\pi}\int_{0}^{\pi} 
Y_{ij}^{*}(\theta, \phi)  |n_\alpha|   \sin\theta \; d\theta d\phi &= I_{ij}^{\alpha},
 \label{eq:ResumingIntegralAppxC}
\end{align} 
the solution of \eqref{eq:AppxCLeastSquaresMinimization} can be written as:
\begin{align}
c_{ij}^{\beta}
=
 \frac{1}{[j_{j}( \mathrm{i} \kappa_{\beta}\xi)]_{\xi=1}'} \sum_{\alpha \in \{x,y,z\}} \chi_{\alpha}^{\beta} I_{ij}^{\alpha}  .
\end{align} 
Alternatively, using again the indices $\ell$ and $m$, expressing the coefficient $\chi_{\alpha}^{\beta}$ as in \eqref{eq:DecoupledHelmholtzEquationCoefficient2a}, and by using the matrix-vector product, the coefficients $c_{\ell m}^{\beta}$ can be more explicitly written as: 
\begin{align}
c_{\ell m}^{\beta} 
=
\frac{k_s \mathbf{g}_{\beta} \cdot \operatorname{diag}\left[I_{\ell m}^{x}, I_{\ell m}^{y} , I_{\ell m}^{z} \right] \cdot \mathbf{m}_0 }{[j_{\ell}( \mathrm{i} \kappa_{\beta}\xi)]_{\xi=1}'} . \label{eq:AppxSolutionOfTheBoundaryValueProblem}
\end{align} 
For some low-orders of $\ell$ ans $m$, the exact solutions of the integrals $I_{\ell m}^{\alpha}$ are presented in Appendix~\ref{sec:AppendixIntegralSolutions}.

From \eqref{eq:AppxSolutionOfTheBoundaryValueProblem} several conclusions can be drawn. First, the zero-order term ($\ell=m=0$) in \eqref{eq:HelmholtzFundamentalSolution} vanishes, which can easily be shown by rewriting the coefficient $c_{00}^{\beta}$ as:
\begin{align}
c_{00}^{\beta} = \frac{k_s \sqrt{\pi} ( \mathbf{g}_{\beta} \cdot  \mathbf{m}_0)}{[j_{0}( \mathrm{i} \kappa_{\beta}\xi)]_{\xi=1}'} = 0, 
\end{align}
which is due to the orthogonality of $\mathbf{m}_0$ and $\mathbf{g}_{\beta}$ with $\beta\in\{1,2\}$ [see \eqref{eq:UnitVectorBase_g2}]. Moreover, we see that
\begin{align}
\psi_{\beta}(\xi=0, \theta, \phi) = 0, 
\end{align} 
which is a consequence of the behavior of the spherical Bessel functions of the first kind at the origin. Since the $j_\ell$ with $\ell\ge1$ are all vanishing at the origin $\xi = 0$, this implies that the total $\psi_{\beta}$ also vanishes at the origin. Note that $j_0(0)=1$, but does not contribute to $\psi_{\beta}$ due to $c_{00}^{\beta} = 0$. From the physical point of view this make sense, since the spin misalignment is caused by the N\'{e}el surface anisotropy and thus, due to symmetry reasons, there is no spin disorder at the center of the spherical NM; the highest misalignment is found at its surface. Second, in the table of Appendix~\ref{sec:AppendixIntegralSolutions} it is seen that the coefficients $I_{\ell, m}^{\alpha}$ vanish for odd $\ell$ or $m$, and that
$I_{\ell, m}^{\alpha} = I_{\ell, -m}^{\alpha}$, so that the expansion coefficients also exhibit this symmetry
\begin{align}
c_{\ell, m}^{\beta} = c_{\ell, -m}^{\beta} .
\end{align}
Taking these properties into account, one can express the solution \eqref{eq:HelmholtzFundamentalSolution} more conveniently in terms of the associated Legendre polynomials $P_{\ell}^{m}(\cos\theta)$ with $\ell=2\nu$ and $m=2\mu$ [note that we use the convention that $Y_{\ell m}(\theta, \phi) = N_{\ell m} P_{\ell}^{m}(\cos\theta) \mathrm{e}^{\mathrm{i}m\phi}$ (p.~378 (14.30.1) in \cite{olver2010nist})]:
\begin{align}
\psi_{\beta}  &=
 \sum_{\nu=1}^{\infty} \sum_{\mu=0}^{\nu} a_{\nu \mu}^{\beta} \Upsilon_{\nu}(\kappa_{\beta}\xi) P_{2\nu}^{2 \mu}(\cos\theta) \cos(2 \mu\phi),  \label{eq:HelmholtzFundamentalSolutionSymmetryA}
\end{align}
where $\Upsilon_{\nu}(\tau)$ was defined in \eqref{eq:SphericalBesselUpsilon} and $a_{\nu \mu}^{\beta}$ in \eqref{eq:CoefficientsHelmholtzEquation}.

The infinite series in \eqref{eq:CoefficientsHelmholtzEquation} is a consequence of the relation between the spherical Bessel functions of the first kind and the ordinary Bessel functions of the first kind~(p.~262 (10.47.3) in \cite{olver2010nist})
\begin{align}
    j_{l}(\tau) = \sqrt{\frac{\pi}{2 \tau }} J_{l + \frac{1}{2}}(\tau), \label{eq:SphericalBesselOrdinaryBesselAA}
\end{align}
and the well known series representation (p.~262 (10.47.3) in \cite{olver2010nist})
\begin{align}
    J_{l}(\tau) = \sum_{s=0}^{\infty} \frac{(-1)^{s} (\tau/2)^{2s+l}}{s!\Gamma(l + s + 1)} . \label{eq:OrdinaryBesselFunctionSeriesAAAA}
\end{align}

\section{Integral Coefficients}\label{sec:AppendixIntegralSolutions}

The integrals $I_{\ell m}^{\alpha}$ can be simplified, since both $n_{\alpha}$ and $Y_{\ell m}^{\ast}(\theta,\phi)$ are separable functions in $\theta$ and $\phi$. By rewriting the spherical harmonics in terms of complex exponentials and associated Legendre polynomials and by using the definition of the Cartesian components of the surface normal vector $\mathbf{n}$ from \eqref{eq:SurfaceNormalVector} [note that we use the convention that $Y_{\ell m}(\theta, \phi) = N_{\ell m} P_{\ell}^{m}(\cos\theta) \mathrm{e}^{\mathrm{i}m\phi}$ (p.~378 (14.30.1) in \cite{olver2010nist}), the integrals are expressed as follows:
\begin{align}
 I_{\ell m}^{x} &= N_{\ell m} K_{\ell m}^{(1)}U_{m}^{(1)} ,
 \\
 I_{\ell m}^{y} &= N_{\ell m}  K_{\ell m}^{(1)} U_{m}^{(2)} ,
 \\
 I_{\ell m}^{z} &= N_{\ell m}K_{\ell m}^{(2)} U_{m}^{(3)} ,
\end{align}
where
\begin{align}
    K_{\ell m}^{(1)} &= \int_{0}^{\pi} P_{\ell }^{m} (\cos\theta) |\sin\theta| \sin\theta d\theta,
    \\
    K_{\ell m}^{(2)} &= \int_{0}^{\pi} P_{\ell }^{m} (\cos\theta)|\cos\theta| \sin\theta d\theta,
    \\
    U_{m}^{(1)} &=  \int_{0}^{2\pi}  \mathrm{e}^{-\mathrm{i}m\phi}|\cos\phi |  \;    d\phi 
    = (-1)^{|m|/2} \frac{2(1 + (-1)^{|m|})}{1 - m^2 + \delta_{|m|,1}},
    \\
    U_{m}^{(2)} &= \int_{0}^{2\pi}  \mathrm{e}^{-\mathrm{i}m\phi}|\sin\phi|   \;   d\phi  
    = \frac{2(1 + (-1)^{|m|})}{1 - m^2 + \delta_{|m|,1}},
    \\
    U_{m}^{(3)} &= \int_{0}^{2\pi}  \mathrm{e}^{-\mathrm{i}m\phi} \;   d\phi = 2\pi \delta_{0, m} ,
    \\
    N_{\ell m} &= \sqrt{\frac{2\ell + 1}{4\pi} \frac{(\ell-m)!}{(\ell +m)!}}.  
\end{align}
The integrals $U_{m}^{(1)}$, $U_{m}^{(2)}$, and $U_{m}^{(3)}$ are solvable straightforwardly by using Euler's formula for the complex exponential and by splitting the region of integration according to the absolute values of the trigonometric functions. In the denominator of $U_{m}^{(1)}$ and $U_{m}^{(2)}$ we included the Kronecker delta $\delta_{|m|,1}$ to take account of the cases $m = \pm 1$.
It is common to express the integrals $K_{\ell m}^{(1)}$ and $K_{\ell m}^{(2)}$ by the substitution $x = \cos\theta$ ($d x = - \sin\theta d\theta$):
\begin{align}
    K_{\ell m}^{(1)} &= \int_{-1}^{1} P_{\ell }^{m} (x) \sqrt{1-x^2} \; d x,
    \\
    K_{\ell m}^{(2)} &= \int_{-1}^{1} P_{\ell }^{m} (x) |x| \; d x .
\end{align}
Using $U_{m}^{(3)} = 2\pi \delta_{0, m}$, we need only to compute $K_{\ell,0}^{(2)}$ such that the associated Legendre polynomials in $K_{\ell,m}^{(2)}$ are reduced to the Legendre polynomials [with one index only (p.~352 in \cite{olver2010nist})]. By considering the symmetry properties of the Legendre polynomials it becomes clear that the integral must vanish for odd $\ell$ and can be simplified for even $\ell$ in the following way:
\begin{align}
    K_{\ell,0}^{(2)} &= (1 + (-1)^{\ell}) \int_{0}^{1} P_{\ell} (x) x \; d x .
\end{align}
The closed-form of $K_{\ell,0}^{(2)}$ is then found in terms of the Gamma function (p.~771 (7.126.1) in \cite{gradshteyn2007table}):
\begin{align}
    K_{\ell,0}^{(2)} &=  \frac{\sqrt{\pi}(1 + (-1)^{\ell})}{4\Gamma(3/2 - \ell/2)\Gamma(\ell/2 + 2)} .
\end{align}
The overall solution for $I_{\ell m}^{z}$ is then written as:
\begin{align}
    I_{\ell m}^{z} =  
    \frac{ \pi(1 + (-1)^{\ell})\sqrt{2\ell + 1}}{4\Gamma(3/2 - \ell/2)\Gamma(\ell/2 + 2)} \delta_{0, m}.
\end{align}
Since we only have to calculate $K_{\ell m}^{(1)} $ for even $m$ (since $U_{m}^{(1)}$ and $U_{m}^{(2)}$ include the term $1 + (-1)^{|m|}$), we may use the index $m = 2\mu$, and by exploiting the parity of the associated Legendre polynomials $P_{\ell}^{m}(-x) = (-1)^{\ell+m}P_{\ell}^{m}(x)$, we find:
\begin{align}
     K_{\ell, 2\mu}^{(1)} &= (1 + (-1)^{\ell}) \int_{0}^{1} P_{\ell }^{2\mu} (x) \sqrt{1-x^2} \; d x . \label{eq:SoltuionKL2Mu}
\end{align}
From these results it is seen that the integrals $I_{\ell m}^{\alpha}$ with $\alpha \in\{x,y,z\}$ vanish for odd $\ell$ and $m$ (note that in \eqref{eq:SoltuionKL2Mu} this is only the case for $m = 2\mu$). For the remaining integrals $K_{2\nu, 2\mu}^{(1)}$ in \eqref{eq:SoltuionKL2Mu}, where $\ell = 2\nu$, we do not give an expression in closed form, but there must exist one in terms of the Gamma function or the Beta function, since~(p.~324 (3.251.2) in \cite{gradshteyn2007table})
\begin{align}
\int_{0}^{1} x^{s} \sqrt{1 - x^2} \; d x = \frac{1}{2} \mathrm{B}\left( \frac{s+1}{2}, \frac{3}{2}\right),
\end{align}
where $\mathrm{B}(\cdot, \cdot)$ is the Beta function (Euler integral), and the associated Legendre functions $P_{2\nu}^{2\mu}$ of even order and degree are true polynomials, as seen for example from the related Rodrigues formula (p.~360 (14.7.14) in \cite{olver2010nist}). Since the remaining integrals are numerically easy to compute, we used Mathematica for the reconstruction of the analytically exact results, which we present in the Table below for some low orders of $\ell$ and $m$.

\begin{table}[!htbp]
\begin{center}
\scalebox{0.7}{
\begin{tabular}{|c||c|c|c|c|c|c|c|}
\hline %%%%%%%%%%%%%%%%%%%%%%%%%%%%%%%%%%%%%%%%%%%
\multicolumn{8}{|c|}{
\parbox[c][1.3cm][c]{22.0cm}{\centering
$\displaystyle I_{\ell m}^{x} = \int_{0}^{2\pi}\int_{0}^{\pi} Y_{\ell m}^{*}(\theta, \phi) |\sin\theta \cos\phi|\sin\theta \; d\theta  d\phi = (-1)^{|m|/2}\frac{2(1 + (-1)^{|m|})(1 + (-1)^{\ell})}{1 - m^2  + \delta_{|m|,1}}  \sqrt{\frac{2\ell + 1}{4\pi} \frac{(\ell-m)!}{(\ell +m)!}} \int_{0}^{1} P_{\ell}^{m}(x) \sqrt{1  - x^2} \; dx$}
} \\
\hline %%%%%%%%%%%%%%%%%%%%%%%%%%%%%%%%%%%%%%%%%%%
\hline %%%%%%%%%%%%%%%%%%%%%%%%%%%%%%%%%%%%%%%%%%%
\diagbox[height=1.1cm]{$\ell$}{$m$}
& 
\parbox[c][1.1cm][c]{3.06cm}{\centering  $ \displaystyle 0 $  }
&
\parbox[c][1.1cm][c]{3.06cm}{\centering  $ \displaystyle  \pm 1 $  }     
&
\parbox[c][1.1cm][c]{3.06cm}{ \centering  $ \displaystyle  \pm 2 $ }
& 
\parbox[c][1.1cm][c]{3.06cm}{\centering   $ \displaystyle   \pm 3$ }       
& 
\parbox[c][1.1cm][c]{3.06cm}{\centering   $ \displaystyle   \pm 4$ }
& 
\parbox[c][1.1cm][c]{3.06cm}{\centering   $ \displaystyle   \pm 5$ }
& 
\parbox[c][1.1cm][c]{3.06cm}{\centering   $ \displaystyle   \pm 6$ }
\\     
\hline %%%%%%%%%%%%%%%%%%%%%%%%%%%%%%%%%%%%%%%%%%%
\hline %%%%%%%%%%%%%%%%%%%%%%%%%%%%%%%%%%%%%%%%%%%
\parbox[c][1.1cm][c]{1.1cm}{\centering  $ \displaystyle  0  $  }
& 
\parbox[c][1.1cm][c]{3.06cm}{\centering  $ \displaystyle \sqrt{\pi} $  }
& 
\parbox[c][1.1cm][c]{3.06cm}{\centering  $ \displaystyle  $  }     
&
\parbox[c][1.1cm][c]{3.06cm}{ \centering  $ \displaystyle  $ }
& 
\parbox[c][1.1cm][c]{3.06cm}{\centering   $ \displaystyle  $ }   
& 
\parbox[c][1.1cm][c]{3.06cm}{\centering   $ \displaystyle  $ }  
& 
\parbox[c][1.1cm][c]{3.06cm}{\centering   $ \displaystyle  $ } 
& 
\parbox[c][1.1cm][c]{3.06cm}{\centering   $ \displaystyle  $ }    
\\      
\hline %%%%%%%%%%%%%%%%%%%%%%%%%%%%%%%%%%%%%%%%%%%
\parbox[c][1.1cm][c]{1.1cm}{\centering  $ \displaystyle 1 $  }
& 
\parbox[c][1.1cm][c]{3.06cm}{\centering  $ \displaystyle 0 $  }
& 
\parbox[c][1.1cm][c]{3.06cm}{\centering  $ \displaystyle  0 $  }     
&
\parbox[c][1.1cm][c]{3.06cm}{ \centering  $ \displaystyle  $ }
& 
\parbox[c][1.1cm][c]{3.06cm}{\centering   $ \displaystyle  $ } 
& 
\parbox[c][1.1cm][c]{3.06cm}{\centering   $ \displaystyle  $ }     
& 
\parbox[c][1.1cm][c]{3.06cm}{\centering   $ \displaystyle  $ }   
& 
\parbox[c][1.1cm][c]{3.06cm}{\centering   $ \displaystyle  $ }       
\\        
\hline %%%%%%%%%%%%%%%%%%%%%%%%%%%%%%%%%%%%%%%%%%%
\parbox[c][1.1cm][c]{1.1cm}{\centering  $ \displaystyle 2 $  }
& 
\parbox[c][1.1cm][c]{3.06cm}{\centering  $ \displaystyle -\frac{\sqrt{ 5 \pi}}{8}$  }
& 
\parbox[c][1.1cm][c]{3.06cm}{\centering  $ \displaystyle  0$  }     
&
\parbox[c][1.1cm][c]{3.06cm}{ \centering  $ \displaystyle  \frac{1}{8} \sqrt{\frac{15\pi}{2}} $ }
& 
\parbox[c][1.1cm][c]{3.06cm}{\centering   $ \displaystyle  $ }   
& 
\parbox[c][1.1cm][c]{3.06cm}{\centering   $ \displaystyle  $ }    
& 
\parbox[c][1.1cm][c]{3.06cm}{\centering   $ \displaystyle  $ } 
& 
\parbox[c][1.1cm][c]{3.06cm}{\centering   $ \displaystyle  $ }        
\\       
\hline %%%%%%%%%%%%%%%%%%%%%%%%%%%%%%%%%%%%%%%%%%%
\parbox[c][1.1cm][c]{1.1cm}{\centering  $ \displaystyle  3 $  }
& 
\parbox[c][1.1cm][c]{3.06cm}{\centering  $ \displaystyle 0$  }
& 
\parbox[c][1.1cm][c]{3.06cm}{\centering  $ \displaystyle  0$  }     
&
\parbox[c][1.1cm][c]{3.06cm}{ \centering  $ \displaystyle 0 $ }
& 
\parbox[c][1.1cm][c]{3.06cm}{\centering   $ \displaystyle 0 $ }   
& 
\parbox[c][1.1cm][c]{3.06cm}{\centering   $ \displaystyle  $ }    
& 
\parbox[c][1.1cm][c]{3.06cm}{\centering   $ \displaystyle  $ }    
& 
\parbox[c][1.1cm][c]{3.06cm}{\centering   $ \displaystyle  $ }     
\\    
\hline %%%%%%%%%%%%%%%%%%%%%%%%%%%%%%%%%%%%%%%%%%%
\parbox[c][1.1cm][c]{1.1cm}{\centering  $ \displaystyle 4 $  }
& 
\parbox[c][1.1cm][c]{3.06cm}{\centering  $ \displaystyle - \frac{3\sqrt{\pi}}{64}$  }
& 
\parbox[c][1.1cm][c]{3.06cm}{\centering  $ \displaystyle  0$  }     
&
\parbox[c][1.1cm][c]{3.06cm}{ \centering  $ \displaystyle \frac{1}{32} \sqrt{\frac{5\pi}{2}} $ }
& 
\parbox[c][1.1cm][c]{3.06cm}{\centering   $ \displaystyle 0 $ }    
& 
\parbox[c][1.1cm][c]{3.06cm}{\centering   $ \displaystyle - \frac{1}{64}\sqrt{\frac{35\pi}{2}} $ }    
& 
\parbox[c][1.1cm][c]{3.06cm}{\centering   $ \displaystyle  $ }   
& 
\parbox[c][1.1cm][c]{3.06cm}{\centering   $ \displaystyle  $ }     
\\          
\hline %%%%%%%%%%%%%%%%%%%%%%%%%%%%%%%%%%%%%%%%%%%
\parbox[c][1.1cm][c]{1.1cm}{\centering  $ \displaystyle  5 $  }
& 
\parbox[c][1.1cm][c]{3.06cm}{\centering  $ \displaystyle 0$  }
& 
\parbox[c][1.1cm][c]{3.06cm}{\centering  $ \displaystyle  0$  }     
&
\parbox[c][1.1cm][c]{3.06cm}{ \centering  $ \displaystyle 0$ }
& 
\parbox[c][1.1cm][c]{3.06cm}{\centering   $ \displaystyle 0 $ }    
& 
\parbox[c][1.1cm][c]{3.06cm}{\centering   $ \displaystyle 0$ }       
& 
\parbox[c][1.1cm][c]{3.06cm}{\centering   $ \displaystyle  0 $ }  
& 
\parbox[c][1.1cm][c]{3.06cm}{\centering   $ \displaystyle  $ }   
\\
\hline %%%%%%%%%%%%%%%%%%%%%%%%%%%%%%%%%%%%%%%%%%%
\parbox[c][1.1cm][c]{1.1cm}{\centering  $ \displaystyle  6 $  }
& 
\parbox[c][1.1cm][c]{3.06cm}{\centering  $ \displaystyle - \frac{5\sqrt{13\pi}}{1024} $  }
& 
\parbox[c][1.1cm][c]{3.06cm}{\centering  $ \displaystyle  0$  }     
&
\parbox[c][1.1cm][c]{3.06cm}{ \centering  $ \displaystyle \frac{\sqrt{1365\pi}}{2048}$ }
& 
\parbox[c][1.1cm][c]{3.06cm}{\centering   $ \displaystyle 0 $ }    
& 
\parbox[c][1.1cm][c]{3.06cm}{\centering   $ \displaystyle -\frac{3}{1024} \sqrt{\frac{91\pi}{2}}$ }       
& 
\parbox[c][1.1cm][c]{3.06cm}{\centering   $ \displaystyle  0 $ }  
& 
\parbox[c][1.1cm][c]{3.06cm}{\centering   $ \displaystyle  \frac{\sqrt{3003\pi}}{2048}$ }   
\\                    
\hline %%%%%%%%%%%%%%%%%%%%%%%%%%%%%%%%%%%%%%%%%%%
\hline %%%%%%%%%%%%%%%%%%%%%%%%%%%%%%%%%%%%%%%%%%%
\multicolumn{8}{|c|}{
\parbox[c][1.4cm][c]{22.0cm}{\centering
$\displaystyle I_{\ell m}^{y} = \int_{0}^{2\pi}\int_{0}^{\pi} Y_{\ell m}^{*}(\theta, \phi) |\sin\theta\sin\phi|\sin\theta \; d\theta d\phi = \frac{2(1 + (-1)^{|m|})(1 + (-1)^{\ell})}{1 - m^2  + \delta_{|m|,1}}  \sqrt{\frac{2\ell + 1}{4\pi} \frac{(\ell-m)!}{(\ell +m)!}}   \int_{0}^{1} P_{\ell}^{m}(x) \sqrt{1  - x^2} \; dx $}
} \\
\hline %%%%%%%%%%%%%%%%%%%%%%%%%%%%%%%%%%%%%%%%%%%
\hline %%%%%%%%%%%%%%%%%%%%%%%%%%%%%%%%%%%%%%%%%%%
\diagbox[height=1.1cm]{$\ell$}{$m$}
& 
\parbox[c][1.1cm][c]{3.06cm}{\centering  $ \displaystyle 0 $  }
&
\parbox[c][1.1cm][c]{3.06cm}{\centering  $ \displaystyle  \pm 1 $  }     
&
\parbox[c][1.1cm][c]{3.06cm}{ \centering  $ \displaystyle  \pm 2 $ }
& 
\parbox[c][1.1cm][c]{3.06cm}{\centering   $ \displaystyle   \pm 3$ }       
& 
\parbox[c][1.1cm][c]{3.06cm}{\centering   $ \displaystyle   \pm 4$ }
& 
\parbox[c][1.1cm][c]{3.06cm}{\centering   $ \displaystyle   \pm 5$ }
& 
\parbox[c][1.1cm][c]{3.06cm}{\centering   $ \displaystyle   \pm 6$ }
\\     
\hline %%%%%%%%%%%%%%%%%%%%%%%%%%%%%%%%%%%%%%%%%%%
\hline %%%%%%%%%%%%%%%%%%%%%%%%%%%%%%%%%%%%%%%%%%%
\parbox[c][1.1cm][c]{1.1cm}{\centering  $ \displaystyle  0  $  }
& 
\parbox[c][1.1cm][c]{3.06cm}{\centering  $ \displaystyle \sqrt{\pi} $  }
& 
\parbox[c][1.1cm][c]{3.06cm}{\centering  $ \displaystyle  $  }     
&
\parbox[c][1.1cm][c]{3.06cm}{ \centering  $ \displaystyle  $ }
& 
\parbox[c][1.1cm][c]{3.06cm}{\centering   $ \displaystyle  $ }   
& 
\parbox[c][1.1cm][c]{3.06cm}{\centering   $ \displaystyle  $ }  
& 
\parbox[c][1.1cm][c]{3.06cm}{\centering   $ \displaystyle  $ } 
& 
\parbox[c][1.1cm][c]{3.06cm}{\centering   $ \displaystyle  $ }    
\\      
\hline %%%%%%%%%%%%%%%%%%%%%%%%%%%%%%%%%%%%%%%%%%%
\parbox[c][1.1cm][c]{1.1cm}{\centering  $ \displaystyle 1 $  }
& 
\parbox[c][1.1cm][c]{3.06cm}{\centering  $ \displaystyle 0 $  }
& 
\parbox[c][1.1cm][c]{3.06cm}{\centering  $ \displaystyle  0 $  }     
&
\parbox[c][1.1cm][c]{3.06cm}{ \centering  $ \displaystyle  $ }
& 
\parbox[c][1.1cm][c]{3.06cm}{\centering   $ \displaystyle  $ } 
& 
\parbox[c][1.1cm][c]{3.06cm}{\centering   $ \displaystyle  $ }     
& 
\parbox[c][1.1cm][c]{3.06cm}{\centering   $ \displaystyle  $ }   
& 
\parbox[c][1.1cm][c]{3.06cm}{\centering   $ \displaystyle  $ }       
\\        
\hline %%%%%%%%%%%%%%%%%%%%%%%%%%%%%%%%%%%%%%%%%%%
\parbox[c][1.1cm][c]{1.1cm}{\centering  $ \displaystyle 2 $  }
& 
\parbox[c][1.1cm][c]{3.06cm}{\centering  $ \displaystyle -\frac{\sqrt{ 5 \pi}}{8}$  }
& 
\parbox[c][1.1cm][c]{3.06cm}{\centering  $ \displaystyle  0$  }     
&
\parbox[c][1.1cm][c]{3.06cm}{ \centering  $ \displaystyle - \frac{1}{8} \sqrt{\frac{15\pi}{2}} $ }
& 
\parbox[c][1.1cm][c]{3.06cm}{\centering   $ \displaystyle  $ }   
& 
\parbox[c][1.1cm][c]{3.06cm}{\centering   $ \displaystyle  $ }    
& 
\parbox[c][1.1cm][c]{3.06cm}{\centering   $ \displaystyle  $ } 
& 
\parbox[c][1.1cm][c]{3.06cm}{\centering   $ \displaystyle  $ }        
\\       
\hline %%%%%%%%%%%%%%%%%%%%%%%%%%%%%%%%%%%%%%%%%%%
\parbox[c][1.1cm][c]{1.1cm}{\centering  $ \displaystyle  3 $  }
& 
\parbox[c][1.1cm][c]{3.06cm}{\centering  $ \displaystyle 0$  }
& 
\parbox[c][1.1cm][c]{3.06cm}{\centering  $ \displaystyle  0$  }     
&
\parbox[c][1.1cm][c]{3.06cm}{ \centering  $ \displaystyle 0 $ }
& 
\parbox[c][1.1cm][c]{3.06cm}{\centering   $ \displaystyle 0 $ }   
& 
\parbox[c][1.1cm][c]{3.06cm}{\centering   $ \displaystyle  $ }    
& 
\parbox[c][1.1cm][c]{3.06cm}{\centering   $ \displaystyle  $ }    
& 
\parbox[c][1.1cm][c]{3.06cm}{\centering   $ \displaystyle  $ }     
\\    
\hline %%%%%%%%%%%%%%%%%%%%%%%%%%%%%%%%%%%%%%%%%%%
\parbox[c][1.1cm][c]{1.1cm}{\centering  $ \displaystyle 4 $  }
& 
\parbox[c][1.1cm][c]{3.06cm}{\centering  $ \displaystyle - \frac{3\sqrt{\pi}}{64}$  }
& 
\parbox[c][1.1cm][c]{3.06cm}{\centering  $ \displaystyle  0$  }     
&
\parbox[c][1.1cm][c]{3.06cm}{ \centering  $ \displaystyle -\frac{1}{32} \sqrt{\frac{5\pi}{2}} $ }
& 
\parbox[c][1.1cm][c]{3.06cm}{\centering   $ \displaystyle 0 $ }    
& 
\parbox[c][1.1cm][c]{3.06cm}{\centering   $ \displaystyle - \frac{1}{64}\sqrt{\frac{35\pi}{2}} $ }    
& 
\parbox[c][1.1cm][c]{3.06cm}{\centering   $ \displaystyle  $ }   
& 
\parbox[c][1.1cm][c]{3.06cm}{\centering   $ \displaystyle  $ }     
\\          
\hline %%%%%%%%%%%%%%%%%%%%%%%%%%%%%%%%%%%%%%%%%%%
\parbox[c][1.1cm][c]{1.1cm}{\centering  $ \displaystyle  5 $  }
& 
\parbox[c][1.1cm][c]{3.06cm}{\centering  $ \displaystyle 0$  }
& 
\parbox[c][1.1cm][c]{3.06cm}{\centering  $ \displaystyle  0$  }     
&
\parbox[c][1.1cm][c]{3.06cm}{ \centering  $ \displaystyle 0$ }
& 
\parbox[c][1.1cm][c]{3.06cm}{\centering   $ \displaystyle 0 $ }    
& 
\parbox[c][1.1cm][c]{3.06cm}{\centering   $ \displaystyle 0$ }       
& 
\parbox[c][1.1cm][c]{3.06cm}{\centering   $ \displaystyle  0 $ }  
& 
\parbox[c][1.1cm][c]{3.06cm}{\centering   $ \displaystyle  $ }   
\\
\hline %%%%%%%%%%%%%%%%%%%%%%%%%%%%%%%%%%%%%%%%%%%
\parbox[c][1.1cm][c]{1.1cm}{\centering  $ \displaystyle  6 $  }
& 
\parbox[c][1.1cm][c]{3.06cm}{\centering  $ \displaystyle - \frac{5 \sqrt{13\pi}}{1024} $  }
& 
\parbox[c][1.1cm][c]{3.06cm}{\centering  $ \displaystyle  0$  }     
&
\parbox[c][1.1cm][c]{3.06cm}{ \centering  $ \displaystyle -\frac{\sqrt{1365\pi}}{2048}$ }
& 
\parbox[c][1.1cm][c]{3.06cm}{\centering   $ \displaystyle 0 $ }    
& 
\parbox[c][1.1cm][c]{3.06cm}{\centering   $ \displaystyle -\frac{3}{1024} \sqrt{\frac{91\pi}{2}}$ }       
& 
\parbox[c][1.1cm][c]{3.06cm}{\centering   $ \displaystyle  0 $ }  
& 
\parbox[c][1.1cm][c]{3.06cm}{\centering   $ \displaystyle  -\frac{\sqrt{3003\pi}}{2048}$ }   
\\                    
\hline %%%%%%%%%%%%%%%%%%%%%%%%%%%%%%%%%%%%%%%%%%%
\hline %%%%%%%%%%%%%%%%%%%%%%%%%%%%%%%%%%%%%%%%%%%
\multicolumn{8}{|c|}{
\parbox[c][1.4cm][c]{22.0cm}{\centering
$\displaystyle I_{\ell m}^{z} = \int_{0}^{2\pi}\int_{0}^{\pi} Y_{\ell m}^{*}(\theta, \phi) |\cos\theta |\sin\theta \; d\theta d\phi =  \frac{\pi (1 + (-1)^{\ell}) \sqrt{2\ell+1}}{4\operatorname{\Gamma}\left( 3/2 - \ell/2\right)
\operatorname{\Gamma}\left( \ell/2 +2 \right) } \delta_{m,0}$}
} \\
\hline %%%%%%%%%%%%%%%%%%%%%%%%%%%%%%%%%%%%%%%%%%%
\hline %%%%%%%%%%%%%%%%%%%%%%%%%%%%%%%%%%%%%%%%%%%
\diagbox[height=1.1cm]{$\ell$}{$m$}
& 
\parbox[c][1.1cm][c]{3.06cm}{\centering  $ \displaystyle 0 $  }
&
\parbox[c][1.1cm][c]{3.06cm}{\centering  $ \displaystyle  \pm 1 $  }     
&
\parbox[c][1.1cm][c]{3.06cm}{ \centering  $ \displaystyle  \pm 2 $ }
& 
\parbox[c][1.1cm][c]{3.06cm}{\centering   $ \displaystyle   \pm 3$ }       
& 
\parbox[c][1.1cm][c]{3.06cm}{\centering   $ \displaystyle   \pm 4$ }
& 
\parbox[c][1.1cm][c]{3.06cm}{\centering   $ \displaystyle   \pm 5$ }
& 
\parbox[c][1.1cm][c]{3.06cm}{\centering   $ \displaystyle   \pm 6$ }
\\     
\hline %%%%%%%%%%%%%%%%%%%%%%%%%%%%%%%%%%%%%%%%%%%
\hline %%%%%%%%%%%%%%%%%%%%%%%%%%%%%%%%%%%%%%%%%%%
\parbox[c][1.1cm][c]{1.1cm}{\centering  $ \displaystyle  0  $  }
& 
\parbox[c][1.1cm][c]{3.06cm}{\centering  $ \displaystyle \sqrt{\pi} $  }
& 
\parbox[c][1.1cm][c]{3.06cm}{\centering  $ \displaystyle  $  }     
&
\parbox[c][1.1cm][c]{3.06cm}{ \centering  $ \displaystyle  $ }
& 
\parbox[c][1.1cm][c]{3.06cm}{\centering   $ \displaystyle  $ }   
& 
\parbox[c][1.1cm][c]{3.06cm}{\centering   $ \displaystyle  $ }  
& 
\parbox[c][1.1cm][c]{3.06cm}{\centering   $ \displaystyle  $ } 
& 
\parbox[c][1.1cm][c]{3.06cm}{\centering   $ \displaystyle  $ }    
\\      
\hline %%%%%%%%%%%%%%%%%%%%%%%%%%%%%%%%%%%%%%%%%%%
\parbox[c][1.1cm][c]{1.1cm}{\centering  $ \displaystyle 1 $  }
& 
\parbox[c][1.1cm][c]{3.06cm}{\centering  $ \displaystyle 0 $  }
& 
\parbox[c][1.1cm][c]{3.06cm}{\centering  $ \displaystyle  0 $  }     
&
\parbox[c][1.1cm][c]{3.06cm}{ \centering  $ \displaystyle  $ }
& 
\parbox[c][1.1cm][c]{3.06cm}{\centering   $ \displaystyle  $ } 
& 
\parbox[c][1.1cm][c]{3.06cm}{\centering   $ \displaystyle  $ }     
& 
\parbox[c][1.1cm][c]{3.06cm}{\centering   $ \displaystyle  $ }   
& 
\parbox[c][1.1cm][c]{3.06cm}{\centering   $ \displaystyle  $ }       
\\        
\hline %%%%%%%%%%%%%%%%%%%%%%%%%%%%%%%%%%%%%%%%%%%
\parbox[c][1.1cm][c]{1.1cm}{\centering  $ \displaystyle 2 $  }
& 
\parbox[c][1.1cm][c]{3.06cm}{\centering  $ \displaystyle \frac{\sqrt{ 5 \pi}}{4}$  }
& 
\parbox[c][1.1cm][c]{3.06cm}{\centering  $ \displaystyle  0$  }     
&
\parbox[c][1.1cm][c]{3.06cm}{ \centering  $ \displaystyle  0$ }
& 
\parbox[c][1.1cm][c]{3.06cm}{\centering   $ \displaystyle  $ }   
& 
\parbox[c][1.1cm][c]{3.06cm}{\centering   $ \displaystyle  $ }    
& 
\parbox[c][1.1cm][c]{3.06cm}{\centering   $ \displaystyle  $ } 
& 
\parbox[c][1.1cm][c]{3.06cm}{\centering   $ \displaystyle  $ }        
\\       
\hline %%%%%%%%%%%%%%%%%%%%%%%%%%%%%%%%%%%%%%%%%%%
\parbox[c][1.1cm][c]{1.1cm}{\centering  $ \displaystyle  3 $  }
& 
\parbox[c][1.1cm][c]{3.06cm}{\centering  $ \displaystyle 0$  }
& 
\parbox[c][1.1cm][c]{3.06cm}{\centering  $ \displaystyle  0$  }     
&
\parbox[c][1.1cm][c]{3.06cm}{ \centering  $ \displaystyle 0 $ }
& 
\parbox[c][1.1cm][c]{3.06cm}{\centering   $ \displaystyle 0 $ }   
& 
\parbox[c][1.1cm][c]{3.06cm}{\centering   $ \displaystyle  $ }    
& 
\parbox[c][1.1cm][c]{3.06cm}{\centering   $ \displaystyle  $ }    
& 
\parbox[c][1.1cm][c]{3.06cm}{\centering   $ \displaystyle  $ }     
\\    
\hline %%%%%%%%%%%%%%%%%%%%%%%%%%%%%%%%%%%%%%%%%%%
\parbox[c][1.1cm][c]{1.1cm}{\centering  $ \displaystyle 4 $  }
& 
\parbox[c][1.1cm][c]{3.06cm}{\centering  $ \displaystyle - \frac{\sqrt{\pi}}{8}$  }
& 
\parbox[c][1.1cm][c]{3.06cm}{\centering  $ \displaystyle  0$  }     
&
\parbox[c][1.1cm][c]{3.06cm}{ \centering  $ \displaystyle 0$ }
& 
\parbox[c][1.1cm][c]{3.06cm}{\centering   $ \displaystyle 0 $ }    
& 
\parbox[c][1.1cm][c]{3.06cm}{\centering   $ \displaystyle 0$ }    
& 
\parbox[c][1.1cm][c]{3.06cm}{\centering   $ \displaystyle  $ }   
& 
\parbox[c][1.1cm][c]{3.06cm}{\centering   $ \displaystyle  $ }     
\\          
\hline %%%%%%%%%%%%%%%%%%%%%%%%%%%%%%%%%%%%%%%%%%%
\parbox[c][1.1cm][c]{1.1cm}{\centering  $ \displaystyle  5 $  }
& 
\parbox[c][1.1cm][c]{3.06cm}{\centering  $ \displaystyle 0$  }
& 
\parbox[c][1.1cm][c]{3.06cm}{\centering  $ \displaystyle  0$  }     
&
\parbox[c][1.1cm][c]{3.06cm}{ \centering  $ \displaystyle 0$ }
& 
\parbox[c][1.1cm][c]{3.06cm}{\centering   $ \displaystyle 0 $ }    
& 
\parbox[c][1.1cm][c]{3.06cm}{\centering   $ \displaystyle 0$ }       
& 
\parbox[c][1.1cm][c]{3.06cm}{\centering   $ \displaystyle  0 $ }  
& 
\parbox[c][1.1cm][c]{3.06cm}{\centering   $ \displaystyle  $ }   
\\
\hline %%%%%%%%%%%%%%%%%%%%%%%%%%%%%%%%%%%%%%%%%%%
\parbox[c][1.1cm][c]{1.1cm}{\centering  $ \displaystyle  6 $  }
& 
\parbox[c][1.1cm][c]{3.06cm}{\centering  $ \displaystyle \frac{\sqrt{13\pi}}{64} $  }
& 
\parbox[c][1.1cm][c]{3.06cm}{\centering  $ \displaystyle  0$  }     
&
\parbox[c][1.1cm][c]{3.06cm}{ \centering  $ \displaystyle 0$ }
& 
\parbox[c][1.1cm][c]{3.06cm}{\centering   $ \displaystyle 0 $ }    
& 
\parbox[c][1.1cm][c]{3.06cm}{\centering   $ \displaystyle 0$ }       
& 
\parbox[c][1.1cm][c]{3.06cm}{\centering   $ \displaystyle  0 $ }  
& 
\parbox[c][1.1cm][c]{3.06cm}{\centering   $ \displaystyle 0$ }   
\\                    
\hline %%%%%%%%%%%%%%%%%%%%%%%%%%%%%%%%%%%%%%%%%%%
\end{tabular}}
\end{center}
\label{table:IntegralTable}
\caption{Values of the integrals \eqref{eq:IntegralCoefficients} for some small values of $\ell$ and $m$.}
\end{table}

\newpage

\section{Derivation of the Fourier Transform of the Magnetization}
\label{sec:Derivation of the Fourier transform}

The Fourier transform of the magnetization vector field $\mathbf{M}(\mathbf{r})$ is written as:
\begin{align}
    \widetilde{\mathbf{M}}(\mathbf{q}) = \frac{1}{(2\pi)^{3/2}} \int_V \mathbf{M}(\mathbf{r})\exp\left( - \mathrm{i}\mathbf{q}\cdot \mathbf{r}\right)d^3r . \label{eq:FourierTransformMagnetizationA1}
\end{align}
In the sequel, we will use dimensionless quantities. For this purpose, we define the dimensionless scattering vector $\boldsymbol{\upsilon} = \mathbf{q} R$, where $R$ is the radius of the nanomagnet, the dimensionless position vector $\boldsymbol{\xi} = \mathbf{r}/R$, and the dimensionless magnetization vector $\mathbf{m} = \mathbf{M}/M_0$, where $M_0$ is the saturation magnetization. Substituting in \eqref{eq:FourierTransformMagnetizationA1} results in
\begin{align}
    \widetilde{\mathbf{M}}(\boldsymbol{\upsilon}) = \frac{R^3M_0}{(2\pi)^{3/2}} \int_V \mathbf{m}(\boldsymbol{\xi})\exp\left( - \mathrm{i}\boldsymbol{\upsilon}\cdot \boldsymbol{\xi}\right)\; d^3\xi.
\end{align}
The dimensionless Fourier transform $\widetilde{\boldsymbol{\mathcal{M}}}$ is then defined as
\begin{align}
    \widetilde{\boldsymbol{\mathcal{M}}}(\boldsymbol{\upsilon}) = \frac{1}{4\pi} \int_V \mathbf{m}(\boldsymbol{\xi})\exp\left( - \mathrm{i}\boldsymbol{\upsilon}\cdot \boldsymbol{\xi}\right) \; d^3\xi, \label{eq:DimensionlessFourierTransformAA}
\end{align}
with
\begin{align}
   \widetilde{\mathbf{M}} = \frac{4\pi R^3M_0}{(2\pi)^{3/2}}\widetilde{\boldsymbol{\mathcal{M}}} .
\end{align}
The next step consists in calculating the Fourier integral of the first-order approximation \eqref{eq:FourierTransformFirstOrderApproximationRealSpace} of the magnetization vector $\mathbf{m}$. Since \eqref{eq:FourierTransformFirstOrderApproximationRealSpace} is formulated in dimensionless spherical coordinates $\xi, \theta, \phi$, it is convenient to express the scattering vector $\boldsymbol{\upsilon}$ in spherical coordinates as well:
\begin{align}
   \boldsymbol{\xi} &= 
   [
   \xi \sin\theta \cos\phi ,
   \xi \sin\theta \sin\phi , 
   \xi \cos\theta 
   ],
    \\
    \boldsymbol{\upsilon} &= 
    [
    \upsilon\sin \theta_q \cos \phi_q , 
    \upsilon \sin\theta_q \sin \phi_q , 
    \upsilon \cos \theta_q 
    ]
    \label{eq:APPX2_qvector} ,
\end{align}
so that the plane-wave expansion of the complex exponential can be used~\cite{jacksonen}
\begin{align}
\exp\left( - \text{i} \boldsymbol{\upsilon} \cdot \boldsymbol{\xi}\right) =  4\pi \sum_{k=0}^{\infty} \sum_{n=-k}^{k} (-\text{i})^{k} j_k(\upsilon \xi) Y_{k n}^{*}(\theta, \phi) Y_{k n}(\theta_q, \phi_q)
\label{eq:APPX2_exponential}.
\end{align}
The Fourier integral \eqref{eq:DimensionlessFourierTransformAA} is then expanded into the following infinite series:
\begin{align}
    &\widetilde{\boldsymbol{\mathcal{M}}}(\boldsymbol{\upsilon}) = 
    \sum_{k=0}^{\infty} \sum_{n=-k}^{k} (-\text{i})^{k}\hat{\boldsymbol{\mathcal{M}}}_{k n}(\upsilon) Y_{k n}(\theta_q, \phi_q), \label{eq:FourierTransformSphericalExpansionA123}
\end{align}
where
\begin{align}
    &\hat{\boldsymbol{\mathcal{M}}}_{k n}(\upsilon) =  \int_{0}^{2\pi}\int_{0}^{\pi}\int_{0}^{1}\mathbf{m}(\boldsymbol{\xi})j_k(\upsilon \xi) Y_{k n}^{*}(\theta, \phi)\xi^2 \sin\theta \; d\xi d\theta d\phi .
   \label{eq:SphericalFourierTransformA}
\end{align}
We now use the infinite series \eqref{eq:HelmholtzFundamentalSolution} for $\psi_{\beta}$ to express the first-order approximation of the magnetization, which leads to
\begin{align}
    \mathbf{m}(\boldsymbol{\xi}) = \mathbf{m}_0 + \sum_{\beta=1}^{2}\mathbf{g}_{\beta} \sum_{\ell=0}^{\infty} \sum_{m=-\ell}^{\ell} c_{\ell m}^{\beta} j_{\ell}(\mathrm{i}\kappa_{\beta}\xi) Y_{\ell m}(\theta,\phi).
    \label{eq:FirstOrderApproximationMagnetizationSphericalHarmonicsA}
\end{align}
Since the integral transform \eqref{eq:SphericalFourierTransformA} is linear, each term in \eqref{eq:FirstOrderApproximationMagnetizationSphericalHarmonicsA} can be separately transformed. For the zero-order term, we obtain
\begin{align}
    \int_{0}^{2\pi}\int_{0}^{\pi}\int_{0}^{1}\mathbf{m}_0 j_k(\upsilon \xi) Y_{k n}^{*}(\theta, \phi)\xi^2 \sin\theta \; d\xi d\theta d\phi   = \sqrt{4\pi} \frac{j_1(\upsilon)}{\upsilon}\mathbf{m}_0  \delta_{k,0} \delta_{n,0} .
\end{align}
This result is well known to the neutron-scattering community as the spherical form factor, corresponding to a uniformly-magnetized spherical particle~\cite{michelsbook}. In the second step, we carry out the integration of the higher-order terms from \eqref{eq:FirstOrderApproximationMagnetizationSphericalHarmonicsA}. The radial and the angular parts in the higher-order terms of \eqref{eq:FirstOrderApproximationMagnetizationSphericalHarmonicsA} are multiplicative, such that the volume integral \eqref{eq:SphericalFourierTransformA} is separable into:
\begin{align}
A_{\ell k}^{\beta}(\upsilon) &= \int_{0}^{1} j_{\ell}(\mathrm{i}\kappa_{\beta}\xi)j_k(\upsilon \xi) \xi^2 d\xi , \label{eq:SphericalFouriertransformSphericalHankeltransform}
\\
B_{\ell m}^{k n}&=
\int_{0}^{2\pi}\int_{0}^{\pi}
  Y_{\ell m}(\theta,\phi)  Y_{k n}^{*}(\theta, \phi) \sin\theta d\theta d\phi . \label{eq:SphericalFouriertransformCoefficientsBlmkn}
\end{align}
As an intermediate result, the integral \eqref{eq:SphericalFourierTransformA} is then rewritten as
\begin{align}
    \hat{\boldsymbol{\mathcal{M}}}_{k n}(\upsilon)&= \sqrt{4\pi} \frac{j_1(\upsilon)}{\upsilon}\mathbf{m}_0 \delta_{k,0} \delta_{n,0} +
    \sum_{\beta=1}^{2}\mathbf{g}_{\beta} \sum_{\ell=0}^{\infty} \sum_{m=-\ell}^{\ell} c_{\ell m}^{\beta}A_{\ell k}^{\beta}(\upsilon) B_{\ell m}^{k n}.
    \label{eq:SphericalFourierTransformIntermediateResult1}
\end{align}
The integral \eqref{eq:SphericalFouriertransformCoefficientsBlmkn} directly corresponds to the orthogonality relation of the spherical harmonics, and thereby we have $B_{\ell m}^{k n} = \delta_{\ell k} \delta_{m n}$~(p.~378 (14.30.8) in \cite{olver2010nist}). Due to the term $\delta_{\ell k}$ in $B_{\ell m}^{k n}$, and since $B_{\ell m}^{k n}$ and $A_{\ell k}^{\beta}$ are multiplicative in \eqref{eq:SphericalFourierTransformIntermediateResult1}, the following spherical Hankel transform results:
\begin{align}
    A_{k k}^{\beta}(\upsilon) &= \int_{0}^{1} j_{k}(\mathrm{i}\kappa_{\beta}\xi)j_{k}(\upsilon \xi) \xi^2 d\xi . \label{eq:SphericalFouriertransformSphericalHankeltransform2}
\end{align} 
In order to calculate these integrals, we replace [using \eqref{eq:SphericalBesselOrdinaryBesselAA}] the spherical Bessel functions of the first kind $j_k(\cdot)$ with the ordinary Bessel functions of the first kind $J_{\iota}(\cdot)$. This yields:
\begin{align}
A_{k k}^{\beta}(\upsilon) =\sqrt{\frac{\pi}{2 \upsilon}} \sqrt{\frac{\pi}{2\mathrm{i} \kappa_{\beta} }}  W_{k}^{\beta}(\upsilon) ,
\end{align}
where
\begin{align}
W_{k}^{\beta}(\upsilon) = \int_{0}^{1}  J_{k + \frac{1}{2}}(\mathrm{i} \kappa_{\beta} \xi)  J_{k + \frac{1}{2}}(\upsilon \xi ) \xi   d \xi .
\end{align}
is the Hommel integral. The result for the integral $W_{k}^{\beta}$ can be found in the textbook by Gradshteyn and Ryzhik (p.~664 (6.521.1) in \cite{gradshteyn2007table}) and is written as
\begin{align}
W_{k}^{\beta}(\upsilon) = - \frac{\upsilon J_{k-\frac{1}{2}}(\upsilon) J_{k + \frac{1}{2}}(\mathrm{i}\kappa_{\beta})-\mathrm{i}\kappa_{\beta} J_{k - \frac{1}{2}}(\mathrm{i}\kappa_{\beta}) J_{k + \frac{1}{2}}(\upsilon)}{\upsilon^2 + \kappa_{\beta}^2} .
\end{align}
Again, upon applying the relation \eqref{eq:SphericalBesselOrdinaryBesselAA}, we write the solution for $A_{kk}^{\beta}(\upsilon)$ more convenient in terms of spherical Bessel functions of the first kind as
\begin{align}
    A_{k k}^{\beta}(\upsilon) &=- \frac{\upsilon j_{k - 1}(\upsilon) j_{k}(\mathrm{i}\kappa_{\beta}) -\mathrm{i}\kappa_{\beta} j_{k - 1}(\mathrm{i}\kappa_{\beta}) j_{k }(\upsilon) }{\upsilon^2 + \kappa_{\beta}^2} .
\end{align}
Now that the integrals are obtained, we have to substitute \eqref{eq:SphericalFourierTransformIntermediateResult1} in \eqref{eq:FourierTransformSphericalExpansionA123}. The term on the first line of \eqref{eq:SphericalFourierTransformIntermediateResult1} only accounts for $k=0$ and $n=0$ in \eqref{eq:FourierTransformSphericalExpansionA123}. Therefore, the contribution of this term to $\widetilde{\boldsymbol{\mathcal{M}}}$ is $j_1(\upsilon)/\upsilon\mathbf{m}_0$, since $Y_{00}(\theta, \phi) = 1/\sqrt{4\pi}$. Since $B_{\ell m}^{k n} = \delta_{\ell k} \delta_{m n}$, the final result is
\begin{align}
    \widetilde{\boldsymbol{\mathcal{M}}}(\boldsymbol{\upsilon}) &= \frac{j_1(\upsilon)}{\upsilon}\mathbf{m}_0  +\sum_{\beta=1}^{2}\mathbf{g}_{\beta} 
    \sum_{k=0}^{\infty} \sum_{n=-k}^{k} (-\text{i})^{k} c_{k n}^{\beta}A_{k k}^{\beta}(\upsilon) Y_{k n}(\theta_q, \phi_q) . \label{eq:FourierTransformSolutionACoeffc}
\end{align}
Using the properties of the coefficients $c_{k n}^{\beta}$ studied in  Appendix~\ref{sec:SolutionOfTheBoundaryValueProblem}, we reformulate \eqref{eq:FourierTransformSolutionACoeffc} in terms of the associated Legendre polynomials, the indices $k = 2\nu$ and $n = 2\mu$, and the coefficients $a_{\nu \mu}^{\beta}$:
\begin{align}
    \widetilde{\boldsymbol{\mathcal{M}}}(\boldsymbol{\upsilon}) = \frac{j_1(\upsilon)}{\upsilon}\mathbf{m}_0  + \sum_{\beta=1}^{2}\mathbf{g}_{\beta} 
    \sum_{\nu=1}^{\infty} \sum_{\mu=0}^{\nu} (-1)^{\nu} a_{\nu \mu}^{\beta} \rho_{\nu}^{\beta}(\upsilon) P_{2\nu}^{2\mu}(\cos \theta_q) \cos(2\mu \phi_q) , \label{eq:FourierTransformSolutionACoeffc2}
\end{align}
where the radial function is given by
\begin{align}
    \rho_{\nu}^{\beta}(\upsilon) = - \frac{\upsilon j_{2\nu- 1}(\upsilon) j_{2\nu}(\mathrm{i}\kappa_{\beta}) -\mathrm{i}\kappa_{\beta} j_{2\nu - 1}(\mathrm{i}\kappa_{\beta}) j_{2\nu }(\upsilon) }{\upsilon^2 + \kappa_{\beta}^2} .
    \label{eq:FourierTransformSolutionRadialFunctionRhoA}
\end{align}
Using once again \eqref{eq:SphericalBesselOrdinaryBesselAA} and the well-known series \eqref{eq:OrdinaryBesselFunctionSeriesAAAA} for the Bessel functions of the first kind, we redefine the spherical Bessel functions of imaginary arguments [as they appear in \eqref{eq:FourierTransformSolutionRadialFunctionRhoA}] as:
\begin{align}
\Upsilon_{\nu}(\tau) &=j_{2\nu}(\mathrm{i}\tau)=   \frac{\sqrt{\pi}}{2} \sum_{s=0}^{\infty} \frac{(-1)^{\nu} (\tau/2)^{2(s+\nu)}}{s!\Gamma(2\nu +s+3/2)},\label{eq:SphericalBesselUpsilonAA}
\\
\mathcal{\aleph}_\nu (\tau) &= \mathrm{i} j_{2\nu-1}(\mathrm{i}\tau) 
=  
\frac{ \sqrt{\pi}}{2} 
\sum_{s=0}^{\infty} \frac{(-1)^{\nu}(\tau/2)^{2(s+\nu)-1}}{s!\Gamma(2\nu + s + 1/2)}    \label{eq:AlephBesselFunctionA},
\end{align}
such that it becomes clear that $\widetilde{\boldsymbol{\mathcal{M}}}(\boldsymbol{\upsilon})$ is a purely real-valued function. The radial function is then rewritten as
\begin{align}
    \rho_{\nu}^{\beta}(\upsilon) = - \frac{\upsilon j_{2\nu- 1}(\upsilon)
    \Upsilon_{\nu}(\kappa_{\beta}) -\kappa_{\beta} \mathcal{\aleph}_\nu (\kappa_{\beta})  j_{2\nu }(\upsilon) }{\upsilon^2 + \kappa_{\beta}^2} .
    \label{eq:FourierTransformSolutionRadialFunctionRhoA2}
\end{align}
By comparing this result to \eqref{eq:SphericalFouriertransformSphericalHankeltransform2}, we find the following pair of spherical Hankel transforms
\begin{align}
    \rho_{\nu}^{\beta}(\upsilon) = \int_{0}^{1}\Upsilon_{\nu}(\kappa_{\beta}\xi) j_{\nu}(\upsilon\xi)\xi^2 d\xi . \label{eq:SphericalHankelTransformUpsilonA22}
\end{align}
By comparing the result \eqref{eq:FourierTransformSolutionACoeffc2} to \eqref{eq:HelmholtzFundamentalSolutionSymmetry} it becomes clear that the angular part is (due to the orthogonality relation of the spherical harmonics) shape invariant under Fourier transformation, while the spherical Hankel transform \eqref{eq:SphericalHankelTransformUpsilonA22} of the radial function $\Upsilon_{\nu}(\kappa_{\beta}\xi)$ remains.

\end{document}